\documentclass[sigconf]{acmart} 

\AtBeginDocument{%
  \providecommand\BibTeX{{%
    \normalfont B\kern-0.5em{\scshape i\kern-0.25em b}\kern-0.8em\TeX}}}

\setcopyright{acmcopyright}
\copyrightyear{2018}
\acmYear{2018}
\acmDOI{10.1145/1122445.1122456}

\usepackage{graphicx}
\usepackage{caption}
\usepackage[skip=0cm,list=true,labelfont=it]{subcaption}
\usepackage[inline]{enumitem}
\usepackage{multirow}
\newcounter{condition}
\newcommand{\nextcondition}{\refstepcounter{condition}\arabic{condition}}

\usepackage{pifont}
\newcommand{\cmark}{\ding{51}}%
\copyrightyear{2023}
\acmYear{2023}
\setcopyright{rightsretained}
\acmConference[CHI '23]{Proceedings of the 2023 CHI Conference on Human Factors in Computing Systems}{April 23--28, 2023}{Hamburg, Germany}
\acmBooktitle{Proceedings of the 2023 CHI Conference on Human Factors in Computing Systems (CHI '23), April 23--28, 2023, Hamburg, Germany}\acmDOI{10.1145/3544548.3581253}
\acmISBN{978-1-4503-9421-5/23/04}

\begin{document}


\title{Overcoming Algorithm Aversion: A Comparison between Process and Outcome Control}

\author{Lingwei Cheng}
\affiliation{%
  \institution{Carnegie Mellon University}
  \city{Pittsburgh}
  \country{USA}}
\email{lingweic@andrew.cmu.edu}

\author{Alexandra Chouldechova}
\affiliation{%
  \institution{Carnegie Mellon University}
  \city{Pittsburgh}
  \country{USA}}
\email{achoulde@andrew.cmu.edu}



\begin{abstract}
Algorithm aversion occurs when humans are reluctant to use algorithms despite their superior performance. Studies show that giving users outcome control by providing agency over how models’ predictions are incorporated into decision-making mitigates algorithm aversion.  We study whether algorithm aversion is mitigated by process control, wherein users can decide what input factors and algorithms to use in model training.  We conduct a replication study of outcome control, and test novel process control study conditions on Amazon Mechanical Turk (MTurk) and Prolific. Our results partly confirm prior findings on the mitigating effects of outcome control, while also forefronting reproducibility challenges.  We find that process control in the form of choosing the training algorithm mitigates algorithm aversion, but changing inputs does not.  Furthermore, giving users both outcome and process control does not reduce algorithm aversion more than outcome or process control alone. This study contributes to design considerations around mitigating algorithm aversion.
\end{abstract}



\maketitle

\section{Introduction}
Recent work on human-AI interaction has closely examined the phenomenon of algorithm aversion, wherein human decision-makers are reluctant to use algorithms even when those algorithms demonstrably outperform expert human judgment \cite{Dawes1989, dawes1979robust, Dietvorst15}. Studies have found that users are more willing to use algorithms as long as they have some control over the outcomes \cite{Dietvorst18}, and are more likely to perceive the algorithms as fair in those settings \cite{Lee19}. This ability to appeal or modify the outcome of a decision once it has been made is termed ``outcome control'' \cite{HOULDEN197813}.  Outcome control can be contrasted with ``process control'', which entails control over the processes that lead to the algorithmic tool (e.g., data curation, the training procedure, etc.).  The effect of various forms of process control on algorithm aversion remains understudied. 

In this work, we explore the effect of process control on algorithm aversion through crowdworker studies conducted on the Amazon Mechanical Turk (MTurk) and Prolific platforms.  We implement outcome control by allowing users to change models' predictions by a restricted amount or freely, and implement process control by allowing users to customize what input factors or model family (e.g., linear regression, trees, etc.) are used in the training process. We aim to understand whether enabling users to directly influence the decision-making process and to express their value judgement through such controls could impact their willingness to use the resulting prediction models, and whether this has any impact on their justice perceptions of those models. At a high level, we ask: \textit{Does process control mitigate algorithm aversion?  Does providing both process control and outcome control more greatly mitigate algorithm aversion than either form of control on its own?}

We answer these questions using data collected from three sets of randomized experiments where we invite crowdworker participants to take the perspectives of end-users of an algorithmic decision-support tool that predicts students' test performance. We first conduct a replication of the original \textit{outcome control} study of \citet{Dietvorst18}, which also relied on a test score prediction task.  Our replication experiment confirms\footnote{When we re-ran the baseline condition of no outcome control as the baseline condition in our second study we found statistically significant differences in algorithm take-up rates for the same condition across experimental runs. We discuss this in further detail in the Results and Discussion sections of the paper.} that allowing participants to change models' predictions significantly increases their likelihood of choosing to use the model, reduces deviation from the models' predictions, and lowers prediction errors.  Interestingly, in later experiments, we find that placing limits on the ways in which participants can change the models’ predictions increases their likelihood of using it, which contradicts a key finding of \cite{Dietvorst18}.  

In our main experiments on process control, we find that not all customization impacts algorithm aversion similarly.  If participants are able to choose the training procedure (i.e., select the structural form of the model), they are more likely to use the models, deviate less from the models' predictions, and commit lower prediction errors, compared to if they have neither process nor outcome control. However, being able to change only which input factors the model considers does not affect the likelihood of model take-up or other measures of reliance and performance.  

When comparing the effects of process vs. outcome control, we find that allowing participants to choose which \textit{training procedure} to use increases the likelihood that users will choose to use the model by the same degree as does allowing participants to change the models' predictions by a limited amount. We also find that providing both forms of control simultaneously \textit{does not} result in further mitigation than providing either outcome or process control alone.  

Additionally, we find no effect of process control on perceived model fairness across study conditions. However, we consistently observe that participants deem the use of models to be less fair as the stakes of the decision become higher. 

Our study contributes to filling an important gap in the empirical investigation of how process control affects algorithm aversion, and compares and contrasts the effects with prior literature on outcome control. The study contributes to an understanding of factors influencing algorithm aversion by showing that users are more likely to use the model if they have some control over the design process. More importantly, we show the effect of process control is more nuanced than previously thought and it depends on the characteristics of the interaction (i.e. the specific type of design choice available to the participants). 

Lastly, we note that our experimental findings are in several cases qualitatively and quantitatively at odds across platforms and experimental runs.  This makes it challenging at times to cleanly present the findings and conclusions, because the results depend on which experiments are compared and across (or within) which platforms.  Having run sets of experiments on both MTurk and Prolific, our study speaks to the challenges of replication and serves as a further cautionary tale for crowdworker studies of human-AI interaction.  Given how many studies of human-AI interaction rely on experiments conducted on crowdworker platforms, we feel it is important to report on reproducibility failures.  Our work thus also contributes to an evolving understanding of reproducibility and platform quality in crowdworker studies (see e.g., \cite{qarout2019platform, 10.1145/3479531, xia2022tragedy}).

\section{Related Work} 
\label{section:related_work} 
Algorithm aversion describes the phenomenon wherein human decision-makers are unwilling to use data-driven algorithms despite being presented with evidence that the algorithms consistently outperform expert human judgment\cite{Dietvorst15}. The degree of aversion has been found to depend on various factors including the characteristics of the tasks \cite{Castelo2019}, the decision makers \cite{Logg2017TheoryOM}, and the interactions between humans and algorithms \cite{Prahl2017}.

Given the many participatory design frameworks---an expansive form of process control---that have been proposed to improve decision-making processes or products in domains including healthcare \cite{Jordan06, estiri16}, recommendation systems \cite{herrmanny2021towards, loepp2015blended}, information technology \cite{Gorski20}, natural language processing \cite{https://doi.org/10.48550/arxiv.2103.04044, godbole2004discriminative}, and public services \cite{Herian12, Saxena20, 10.1145/3442188.3445938}, it is important to better understand the relationship between process control and algorithm take-up.  While there is now an increasing emphasis on the use of participatory design methods for AI development, algorithms mostly continue to be designed by third-party researchers and organizations that may not fully understand users' needs and values.  Proponents of participatory methods argue that current practices limit the utility, usability, and take-up of the algorithms. In various applications of algorithms such as child welfare \cite{10.1145/3313831.3376229, Bosk18, Saxena20} and health \cite{Day_Humphrey_Cockcroft_2017, estiri16, Qian16}, users repeatedly report that they are unsatisfied or feel discouraged from using algorithms because they could not influence what and how evidence is used or should be interpreted. In a child maltreatment screening case study surveying four agencies \cite{Bosk18}, caseworkers diverge from a decision support tool's recommendation because they assess the meaning of demographic factors and caregivers' history differently. In our study, we do not study domain-specific applications, but rather investigate specific elements of the design process by allowing participants to specify what input factors and training procedures (i.e., model hypothesis class/structure) the models can consider, which specifically address some of the concerns mentioned above. Both interventions are proposed by \cite{herrmanny2021towards} as part of the ``engage'' design strategy for integrating users in system designs. Our study will allow us to critically assess the effects of such proposed interventions.

We form our hypotheses around the effects of process control based on recent systematic reviews of the algorithm aversion literature. \citet{Burton20} categorizes the causes of algorithm aversion into five themes: (i) false expectations of algorithm decisions by decision-makers; (ii) decision-makers' lack of decision autonomy; (iii) the lack of social or economic incentives for decision-makers to use algorithms; (iv) misalignment between human and algorithmic decision-making processes; and (v) conflicting concepts of rationality between decision-makers who often seek to minimize uncertainty and algorithms that seek to reduce risks. Correspondingly, algorithm literacy, human-in-the-loop decision-making, behavioral design, engaging human intuition, and incorporating ecologically valid decision constraints are proposed as practical solutions to the problem of algorithm aversion. Similarly, \citet{jussupow2020we} conducted an extensive review of experimental studies on algorithm aversion and developed a framework that further shows how algorithm agency, performance, perceived capabilities, human involvement, human agents' expertise and social distance can influence whether users develop algorithm aversion. According to their framework, human involvement in training and using the algorithms can increase perceived ability to perform the task and human agency, both of which can reduce algorithm aversion. Whereas human-in-the-loop (HITL) studies commonly focus on various forms of outcome control or transparency, we explore \textit{process control} effects.  Specifically, we focus on the decision autonomy component which stipulates that ``human decision makers must feel in control and confident enough to place trust in an algorithm to act on its judgement'' \cite{Burton20}.  

Increasing users' perceptions of decision autonomy can mitigate algorithm aversion, even when that autonomy is illusory. \citet{Dietvorst18} shows that participants are more willing to use algorithms as long as they have some control over the outcome, and subsequently perform better on the forecasting task. This effect was found to be largely insensitive to the extent to which users were able to control the outcome (i.e., deviate from the models' predictions), a finding that does not replicate in our data. In another study, users are better at distinguishing misinformation if they could interact with a static checklist \cite{heuer2022comparative}. Furthermore, \citet{Vaccaro18} shows that this feeling of control can either come from real control over the algorithm making process, or from making adjustments to the process that have little or no bearing on the actual functioning of the algorithm. They find that users felt more satisfied with their social media feed when options to change their feed were present even when these controls did not work. This highlights the possibility (and associated risk) that even an illusion of autonomy can mitigate algorithm aversion. In our study we faithfully implement the design changes selected by study participants and do not consider study conditions involving deception.  

Additionally, we hypothesize that process control may mitigate algorithm aversion by improving perceptions of procedural justice. When people perceive they are treated by the decision-makers in a fair and just manner, they are more likely to perceive the outcome as legitimate and are more likely to comply with the decision and cooperate with the decision-makers, irrespective of the outcome \cite{Tyler2015}. Empirically the positive relationship between perceived fairness and outcome compliance have been demonstrated in a variety of fields including employment \cite{Bies93}, criminal justice \cite{Mazerolle13, Simmons2020, Sunshine03}, and local government decisions\cite{Herian12}. 

Importantly, people react similarly to procedural justice whether the decision-maker is a computerized system or a human being\cite{OTTING201827}. Prior studies of procedural justice perceptions in the context of AI systems have shown that process control improves the perceived fairness of systems, while the effect of standards clarity and outcome explanations is less clear \cite{Lee19}. By engaging a small group of stakeholders at a local food rescue organization through a series of in-person meetings to elicit their beliefs, and implement and evaluate machine learning models based on their beliefs, \citet{WeBuildAI} found that participatory design improved participants' perceived procedural fairness, confidence in models' representation of their own beliefs, and distributive outcomes. 
In studying user interactions with social media, \citet{10.1093/jcmc/zmac010} find that allowing users to tinker with what keywords to use in the classification of hate speech enhance users' trust and agreement with the algorithm. In our experiments, users are able to express their value judgements through modifications to the model training process. We measure users' procedural justice perceptions of the models along several dimensions to assess whether they vary with process control.  

Lastly, in contrast to algorithm aversion, a recent study \cite{LOGG201990} has reported on ``algorithm appreciation'', where algorithms are preferred to human experts in decision-making. The authors attempt to bridge the opposing phenomena by calling for future research on what they call ``theory of machine''. This concept considers how people infer the intentions and beliefs of algorithms, similar to the idea of ``theory of mind'' \cite{dennett1987intentional}. \citet{sundar2008main}'s findings on how positive and negative machine heuristics can mediate trust and agreement with algorithms lend further credence to this form of underlying mechanism. More recently \citet{10.1145/3479864} show how framing can lead to different perceived competence or expert power of algorithms vs. human, and lead to either algorithm aversion or appreciation. Viewed in this light, we believe that there could be a set of continuous malleable factors that can be adjusted to influence the attitudes towards algorithms from least to most favorable. Although our study does not directly address this topic, we present a more nuanced picture of the various effects of process control, and demonstrate the effects of controls do not simply ``stack'': the whole is often found to be less than the sum of its parts.

\section{Study Overview and Research Questions}
\label{section:RQs}
Our experiments were conducted over the course of three studies, which were designed to: (i) replicate the study of \citet{Dietvorst18} and establish a baseline effect of outcome control for later comparison with process control; (ii) assess the effect of process control on algorithm aversion; and (iii) assess the joint effect of process and outcome control on algorithm aversion.  

In [Study 1], which we conduct on MTurk, we replicate the work by \citet{Dietvorst18} on mitigating algorithm aversion.\footnote{We also tested a new experimental condition that relied on an alternative bonus scheme to assess whether participant behavior was consistent with loss aversion bias \cite{kahneman1991anomalies}.  Those results are presented in Appendix ~\ref{appendix:adjust_by_10_alt}.}  Given the replication and reproducibility crisis affecting numerous scientific disciplines \cite{baker2015over, fanelli2018science, ioannidis2005most}, it is scientifically valuable to conduct studies seeking to reproduce seminal prior findings.  Research has shown how conducting even a small number of replication studies can effectively weed out false research findings \cite{moonesinghe2007most}.  Furthermore, conducting a replication study in our context allows us to make direct comparisons of the magnitude of the effect of outcome control (if reproduced) and process control (if observed) on mitigating algorithm aversion.

Next, we carry out our main study on process control [Study 2] and its interactions with outcome control [Study 3]. We ran the two studies sequentially: first on MTurk, and then on Prolific as a (self-)replication. Using two platforms allows us to check the robustness of the new interventions. As we discuss when presenting the results, we were surprised by the magnitude of the difference we observed in our results when comparing across batches of MTurk experiments and across the MTurk and Prolific platforms.  

As we discuss in further detail in \textsection\ref{section:participants_tasks_procedures} below, the experiments center on the task of predicting students' reading test scores using data available on students and their parents.  Our primary outcomes include the likelihood of choosing to use the model (i.e., to rely on the models' predictions for determining participants' bonus payouts), performance on prediction tasks, and deviation from the model. The performance is measured as the Average Absolute Error (AAE) which is the average of the absolute differences between participants' predictions and the ground truth student reading test scores. In the case where the participants opted to use the models' predictions and could not modify them, the models' predictions then become the participants' predictions. The deviation from the model is measured as the Average Absolute Deviation (AAD) which is the average of the absolute differences between models' and participants' predictions. It is zero for participants who chose to use the models and could not modify the models' predictions. 

Our secondary outcome variables include perceptions of transparency and fairness of the models, how well the models represent one's views, and confidence in the models' estimates. Participants reported these measures on a five-point Likert scale in the surveys. We also obtained text responses from participants on their justifications for using the models and making certain design choices to enrich our analysis. 

We study the following research questions:
\begin{enumerate}
\item\label{rq:replication}
Are participants who can modify the models' predictions more likely to choose to use the model compared to participants who cannot modify the predictions?  Do they make lower-error estimates?  Are they less likely to deviate from the model's estimates? (Replication study of \citet{Dietvorst18}) [Study 1]

\item\label{rq:design} Are participants who can modify the models' design more likely to choose the model compared to participants who cannot design the model?  Do they make lower-error estimates?  Are they less likely to deviate from the model's estimates? [Study 2]

\item\label{rq:secondary} Do participants who can change the models' design also report having more confidence in the model's estimates, and perceive the model to be more representative of their view, more transparent, and more fair? [Study 2, Secondary outcomes]

\item\label{rq:comparison} How do the effects of process control compare to the effects of outcome control? [Study 3]

\item\label{rq:interaction} Are there any interaction effects between being able to modify models' predictions and designs? If so, are these interaction effects additive or subtractive (i.e. if more design opportunities are strictly better than no design at all or if they could potentially cancel each other out in terms of their effects on our primary outcomes) [Study 3]
\end{enumerate}
Since the study conditions involve variations of similar treatment components, we summarize all the conditions in Table ~\ref{table:study_conditions} for easy comparison.  We indicate when conditions are identical or sufficiently similar to merit direct comparison as replications of the same condition. 
Specifically, we note that conditions \textit{(\ref{condition:cant-change-outcome}) can't-change-outcome and (\ref{condition:cant-design-ur}) can't-design (use restricted) are the exact same conditions}.  Each serves as the baseline condition for its respective study, and it is meaningful to directly compare those conditions across studies and expect the observed outcomes to be similar.  Conditions \textit{(\ref{condition:adjust-by-10}) adjust-by-10} and \textit{(\ref{condition:cant-design-uf}) can't-design (use freely)} differ only in whether participants are able to deviate by at most 10 points (\ref{condition:adjust-by-10}) or arbitrarily from the model's predictions (\ref{condition:cant-design-uf}). While \citet{Dietvorst18} do not directly consider arbitrary deviation, they explore a range of deviation levels and conclude that the level of permitted deviation does not qualitatively affect the results.  Thus, we expect the outcomes in condition (\ref{condition:cant-design-uf}) to be similar to those in (\ref{condition:adjust-by-10}).

\begin{table*}[h]
\caption{All Study Conditions}
\small\centering
\begin{tabular}{p{1.5cm}p{1cm}p{2.1cm}p{1cm}p{1cm}p{1cm}p{1cm}p{1.5cm}}
\centering Study & \centering {Number} & \centering Name &  \centering Choose if Use Model  & \centering Change Model Output by 10 & \centering Change Model Output Freely & \centering Change Inputs & Change (Training) Algorithm \\
\toprule
\multirow{3}{=}{Outcome Control [study 1]}   & 
(\nextcondition)\label{condition:cant-change-outcome}*          &  can't-change-outcome*        & \centering \cmark                                                & \centering $\cdot$                                        & \centering $\cdot$                               & \centering $\cdot$                & \centering $\cdot$    \cr                                        
& (\nextcondition)\label{condition:use-freely}    & use-freely               & \centering $\cdot$                                                 & \centering $\cdot$                                        & \centering \cmark                              & \centering $\cdot$                & \centering $\cdot$   \cr                                  
&(\nextcondition)\label{condition:adjust-by-10}** & adjust-by-10**         & \centering \cmark                                               & \centering \cmark                                       &\centering $\cdot$                               & \centering$\cdot$                & \centering$\cdot$      \cr                                      
\hline
\multirow{3}{=}{Process Control [Study 2]} &  (\nextcondition)\label{condition:cant-design-ur}*  &  can’t-design (use restricted)*    &\centering \cmark                                              & \centering $\cdot$                                        & \centering $\cdot$                               & \centering $\cdot$                & \centering $\cdot$     \cr                                         
& (\nextcondition)\label{condition:change-input-ur}  & change-input (use restricted)   & \centering \cmark                                                & \centering $\cdot$                                        & \centering $\cdot$                               & \centering \cmark               & \centering $\cdot$   \cr                             
& (\nextcondition)\label{condition:change-algorithm-ur} &  change-algorithm (use restricted) & \centering \cmark                                              & \centering $\cdot$                                        & \centering $\cdot$                               & \centering $\cdot$                & \centering \cmark\cr              
\hline
\multirow{3}{=}{Process \& Outcome Control [Study 3]}   & (\nextcondition)\label{condition:cant-design-uf}**          & can’t-design (use freely)**     & \centering \cmark                                                   & \centering $\cdot$                                        & \centering \cmark                              & \centering $\cdot$                & \centering $\cdot$       \cr                              
&  (\nextcondition)\label{condition:change-input-uf}  & change-input (use freely)  & \centering \cmark                                                  & \centering $\cdot$                                        & \centering \cmark                              & \centering \cmark               & \centering $\cdot$          \cr                  
& (\nextcondition)\label{condition:change-algorithm-uf} &  change-algorithm (use freely)    & \centering \cmark                                                 & \centering $\cdot$                                        & \centering \cmark                              & \centering $\cdot$                & \centering \cmark   \cr                                    
\bottomrule   
\multicolumn{8}{l}{\footnotesize * (1) can't-change-outcome and (4) can't-design (use restricted) conditions are the same.}\\
\multicolumn{8}{l}{\footnotesize * They serve as the control group in their respective studies. }\\
\multicolumn{8}{l}{\footnotesize ** (7) can't-design (use freely) is like condition (3) adjust-by-10 but with no limit on the user's deviation from model predictions.}\\
\end{tabular}
\label{table:study_conditions}
\end{table*}

\section{Participants, Tasks, and Procedures}
\label{section:participants_tasks_procedures}
Here we describe the participants, tasks, and procedures for studies conducted on MTurk. The replication on Prolific follows the same procedures with a few platform-specific adaptations that are noted in Appendix ~\ref{appendix:prolific_replication}.

\subsection{Participants}
\label{section:participants_description}
We required participants on MTurk to satisfy the following three criteria: (1) living in the US; (2) having a Human Intelligence Task (HIT) approval rate larger or equal than 97\%; and (3) having completed at least 1000 HITS. We require the participants to live in the US so they would be more familiar with the contexts of the prediction task which involves predicting reading test scores for students from the US. Lastly, participants are only able to participate in the experiments once. The numbers of participants for all study conditions by platforms are summarized in Table~\ref{table:recruitment}, Appendix~\ref{appendix:recruitment}.

We dropped responses that were incomplete or took more than three hours to complete and where participants spent less than one second per prediction task. The average time spent on the entire survey is 19.6 minutes with a standard deviation of 9.5 minutes across all three studies. The average time spent on the prediction task is 7.1 minutes with a standard deviation of 4.3 minutes. This represents a reasonable time range we expect for participants to complete the tasks. 

Additionally we designed two types of attention check questions. One requires participants to manually type a statement in order to advance. All participants who completed the surveys passed this attention check. The second type of check involved two questions presented at the beginning and at the end of the survey. The task entails that participants predict students' \textit{percentile} reading scores, which are by definition numbers between 1 and 100.  At the beginning of the survey, participants were provided with the definition of percentiles and accompanying examples, and were asked to select correct statements about the meaning of percentile scores among several listed options. They were asked again to answer a true or false question in the end of the survey relating to percentile scores. Not all participants passed the attention check questions. However, we found no statistically significant difference in algorithm aversion when comparing the data on participants who passed both attention checks compared to those who did not ($\chi^2(1, N=2,527) = 2.33, p = 0.13$ for testing differences in the likelihood of algorithm take-up). Since filtering participants on the attention check reduces sample size and power without qualitatively affecting the results, in the main text we present findings on the full data.  We provide results on the subset of participants who passed both attention checks in Appendix~\ref{appendix:filtered_results}, which are qualitatively identical to the findings we present here.

Table ~\ref{table:demographics_all_unfiltered} summarizes the characteristics of participants for MTurk and Prolific. Compared to the demographics of the US population, the MTurk participant sample was disproportionately white, male, non-Hispanic, and highly-educated.  These observed demographics are typical of MTurk, as documented in prior studies \cite{berinsky_huber_lenz_2012, casey2017intertemporal}. Interestingly on Prolific, the sample population had a higher percentage of female participants (65.7\%) compared to  MTurk (40.6\%). Prolific participants were also younger with an average age of 34.6 compared to 38.3 on MTurk. Prolific users tended to have slightly less confidence in math and had done fewer studies previously related to algorithms on the platform. Overall, Table ~\ref{table:demographics_all_unfiltered} shows not only are crowdworkers different from the general population, but they are also different across platforms. 

\begin{table}[h]\centering\small
\def\sym#1{\ifmmode^{#1}\else\(^{#1}\)\fi}
\caption{Demographics of Participants from MTurk, Prolific and Both (Pooled)}
\begin{tabular}{p{5.8cm}p{0.6cm}p{0.6cm}p{0.6cm}}
& Mturk & Prolific & Pooled \\
\hline\hline
\% Female                                                               & 40.6  & 65.7     & 51.2   \\ [0.02cm]
\% Male                                                                & 56.8  & 28.0     & 43.5   \\ [0.02cm]
\% White                                                                 & 84.6  & 74.4     & 79.9   \\ [0.02cm]
\% Black or African American                                             & 7.7   & 6.9      & 7.3    \\ [0.02cm]
\% Hispanic or Latino                                                    & 10.9  & 10.8     & 10.8   \\ [0.02cm]
Average Age (Year)                                                           & 38.3  & 34.6     & 36.6   \\ [0.02cm]
\% Some college and above                                                & 92.2  & 87.4     & 90.0   \\ [0.02cm]
\% High school/GED                                                       & 7.5   & 11.4     & 9.3    \\ [0.02cm]
Average Confidence in Math            & 3.3  & 2.6     & 3.0   \\ [0.02cm]
\% Participated in Algorithm-related Studies Before  & 78.2  & 51.0       & 65.6     \\ [0.02cm]
\hline
N   & 1,355  & 1,172     & 2,527 \\
\hline\hline
\multicolumn{4}{l}{\footnotesize This sample does not include adjust-by-10 (proposed bonus scheme) described in }\\
\multicolumn{4}{l}{\footnotesize Appendix~\ref{appendix:adjust_by_10_alt}.}\\
\multicolumn{4}{l}{\footnotesize Response categories where few respondents selected or declined to answer are dropped.}\\
\multicolumn{4}{l}{\footnotesize Average confidence in math is converted from a survey question using a 5-point }\\
\multicolumn{4}{l}{\footnotesize Likert scale where 1 is "not confident" and 5 is "extremely confident."}\\
\end{tabular}
\label{table:demographics_all_unfiltered}
\end{table}

\subsection{Procedures} \label{sec:procedures}
The studies were administered as interactive online surveys. Participants began by giving consent and entering their MTurk ID. They would then read a short paragraph giving examples of what \textit{percentile scores} mean and answer a multiple choice attention check question to show they have correctly understood the concept.

\subsubsection{Randomization}
For each study, participants were randomized according to Table~\ref{table:study_conditions}.

\subsubsection{Process Control Interventions}
\label{section:process_control_interventions}
All participants were informed that their task was to predict reading test \textit{percentile scores} between $[0,100]$ for 20 high school sophomores. The task was constructed using publicly available Programme for International Student Assessment (PISA) test data from 2009 \cite{mit_opencourseware} on students from the US. Each student is associated with 22 variables falling into five categories: student demographics, English study experience, school conditions, parental characteristics such as parent's education level and employment status, and family characteristics such as if each of the family members was born in the US. Participants were assured that the data points are from real students. Depending on the condition to which they were randomized, participants then followed one of the three paths below. 

\paragraph{Cannot Design Models}
For participants who cannot design the model, namely participants in the can't-change-outcome, use-freely, adjust-by-10, can't-design (use freely), can't-design (use restricted) conditions, they were told that they would have access to all the information on the students and, if they so chose, predictions produced by a statistical model. They were told that the model is based on data from thousands of high school sophomores and it uses the same variables they saw. They were told that across all students, the model's prediction is off by 19.7 percentiles on average, although it may perform better or worse for the small set of students that they would be asked about in this experiment. 

\begin{figure*}[h]
  \begin{minipage}[c]{0.3\textwidth}
    \includegraphics[width=\textwidth]{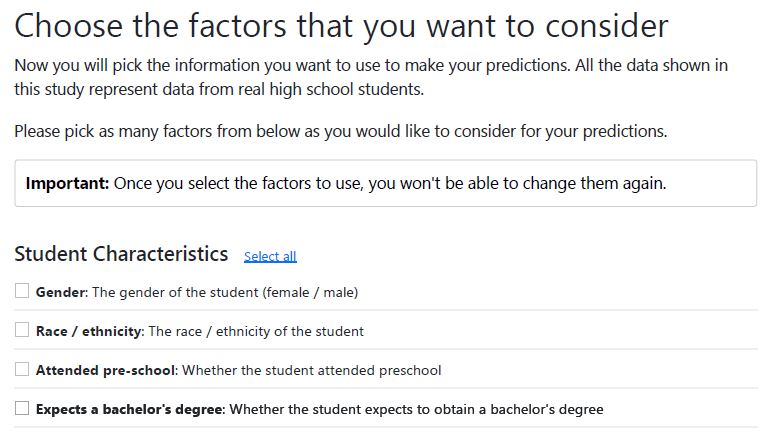}
  \end{minipage}
  \begin{minipage}[c]{0.3\textwidth}
  \centering
    \includegraphics[width=\textwidth]{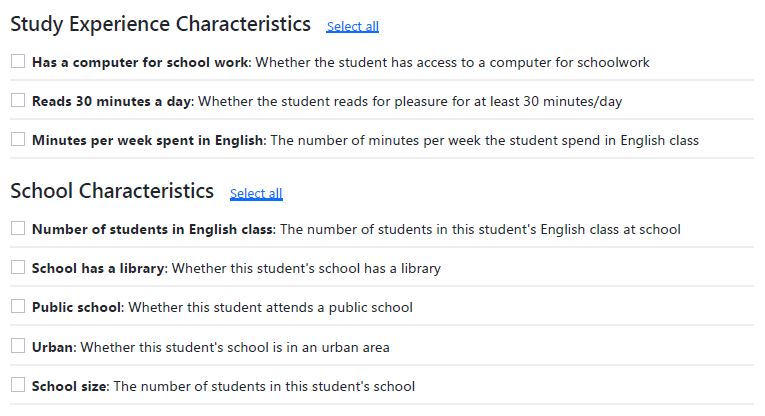}
  \end{minipage}
    \begin{minipage}[c]{0.35\textwidth}
  \centering
    \includegraphics[width=\textwidth]{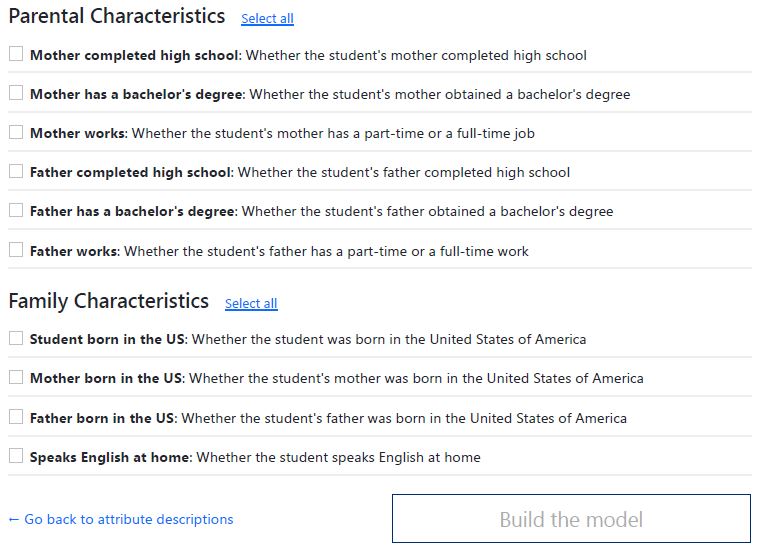}
  \end{minipage}
\caption{Interface where Participants Chose Input Factors for the Models. These screenshots show the single page (broken down into three screenshots here) where participants were presented with all variables associated with students. The information falls into five categories including characteristics of the students, their study experience, school, parents, and family. Each variable is accompanied by a brief explanation of what it is and a check mark if the participants wanted to use the variable. At the end of the page, the participants hit the "Build the model" button and proceed to the build page. }
\Description{Interface where Participants Chose Input Factors for the Models. These screenshots show the single page (broken down into three screenshots here) where participants were presented with all variables associated with students. The information falls into five categories including characteristics of the students, their study experience, school, parents, and family. Each variable is accompanied by a brief explanation of what it is and a check mark if the participants wanted to use the variable. At the end of the page, the participants hit the "Build the model" button and proceed to the build page.}
\label{fig:chose_input}
\end{figure*}

\paragraph{Change Model Inputs}
For participants in the change-input (use freely) and change-input (use restricted) conditions, they learned that they would have the opportunity to build a statistical model by specifying the input factors in the model. They were told that they would see the model's performance first and decide if they want to use it for the task. Next, they would read about the students' data and select variables for the models to use. Figure ~\ref{fig:chose_input} illustrates this interface. 

\begin{figure}[h]
\includegraphics[width=0.5\textwidth]{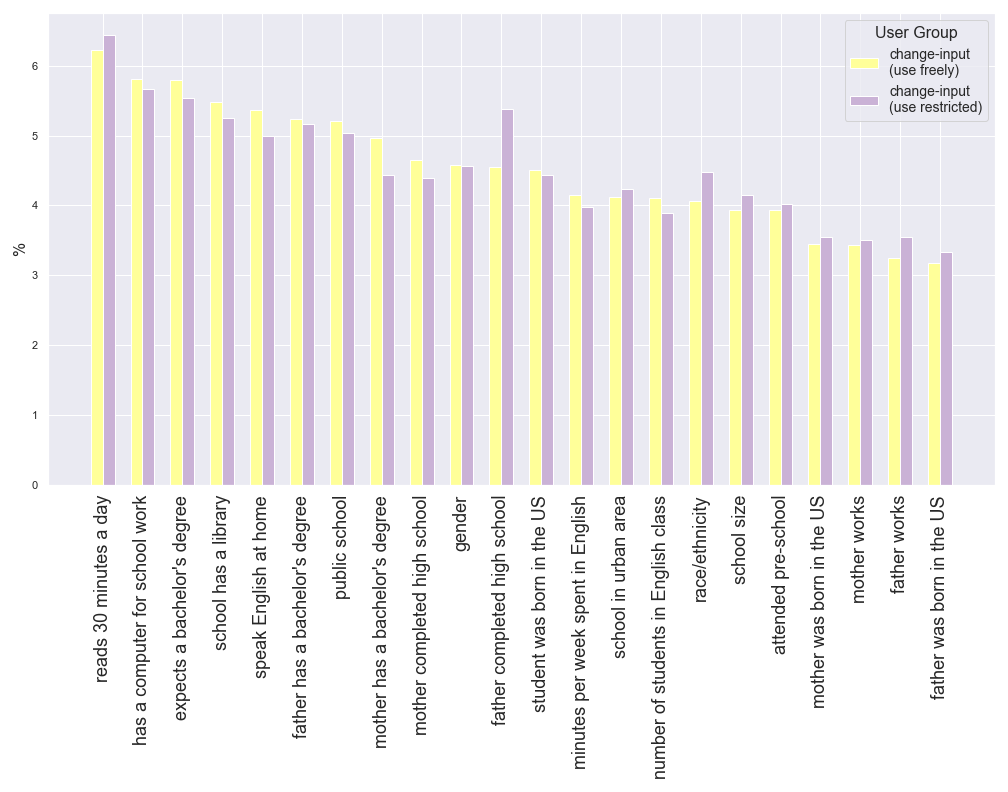}
\caption{Input Factors Selected by Participants. Each bar corresponds to the percentage of the times the variable has been selected out of all the variables. It is normalized so that all bars for each user group sum up to one. The most popular variables include if the student reads 30 minutes a day, has a computer for school work, and if they expect a bachelor's degree. Interestingly demographic features such as gender and race/ethnicity were selected by about or below average rate.}
\Description{Input Factors Selected by Participants. Each bar corresponds to the percentage of the times the variable has been selected out of all the variables. It is normalized so that all bars for each user group sum up to one. The most popular variables include if the student reads 30 minutes a day, has a computer for school work, and if they expect a bachelor's degree. Interestingly demographic features such as gender and race/ethnicity were selected by about or below average rate.}
\label{fig:vars_selected}
\end{figure}

While it may seem obvious to algorithm experts that one should simply choose to use as much information as possible, we find that only 17\% of participants who can change input used all the features. Figure ~\ref{fig:vars_selected} shows the distribution of features chosen by the change-input groups. Participants more frequently chose features directly related to students' reading experience such as if they read at least 30 minutes instead of their gender or race/ethnicity. The text responses suggest most participants selected variables based on their particular assumptions of how the variables should be related to test outcomes. A minority of the participants also expressed that they would not choose demographic variables or variables that are outside of the control of the students to avoid being biased. For example, one participant said that "honestly, even though I knew picking more factors could possibly make the model more accurate, I just wanted to be fair and didn't want to discriminate in any way."

After they selected the variables, they would wait for 15 seconds while a linear regression model was being constructed using the selected variables as input.  This delay was built in to create the impression that model training entailed some time rather than being a computationally instantaneous process.  

\paragraph{Change Model Algorithms}
For participants in the change-algorithm (use freely) and change-algorithm (use restricted) conditions, they similarly learned about the students' data and that they would have the opportunity to build a statistical model by choosing the type of learning algorithm used in training. They were told that they would see the model's performance first and decide if they want to use it for the task. Next, the participants would learn about the training algorithms they can choose, which included linear regression, lasso, trees, random forests, and K-nearest neighbors algorithms. 

\begin{figure*}[h]
  \begin{minipage}[c]{0.3\textwidth}
    \includegraphics[width=\textwidth]{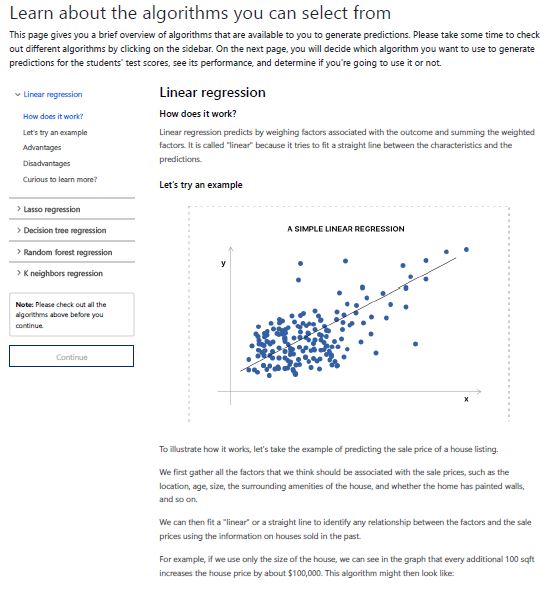}
  \end{minipage}
  \begin{minipage}[c]{0.3\textwidth}
  \centering
    \includegraphics[width=\textwidth]{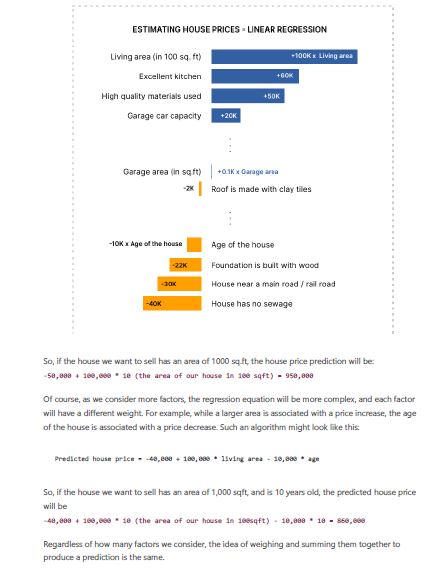}
  \end{minipage}
    \begin{minipage}[c]{0.35\textwidth}
  \centering
    \includegraphics[width=\textwidth]{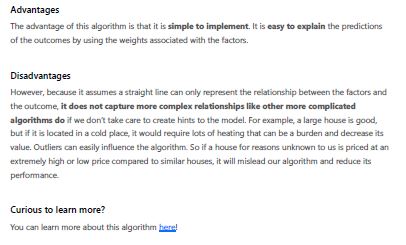}
  \end{minipage}
\caption{Interface where Participants Chose Algorithms for the Models. These screenshots show the single page (broken down into three screenshots here) where participants were given explanations of all algorithms. Participants can choose to view linear regression, lasso, decision trees, random forest, and K-nearest neighbors algorithms on the left navigation bar. Each algorithm explanation is broken down into: (1) How does it work: a brief intuitive explanation of the algorithm; (2) Let's try an example: an application of the algorithm on Ames Housing data; (3) Advantages: any advantage of the algorithm such as if it is generally easy to understand and implement or accurate; (4) Disadvantages: any drawbacks of the algorithm such as the difficulty associated with explaining its results or failures to capture the more complex relationship between the predictors and the outcome; (5) Curious to learn more: a link to the Wikipedia page of the algorithm. Participants need to click through all the algorithms before they could continue to the next page where they decided on which algorithm to use. }
\Description{Interface where Participants Chose Algorithms for the Models. These screenshots show the single page (broken down into three screenshots here) where participants were given explanations of all algorithms. Participants can choose to view linear regression, lasso, decision trees, random forest, and K-nearest neighbors algorithms on the left navigation bar. Each algorithm explanation is broken down into: (1) How does it work: a brief intuitive explanation of the algorithm; (2) Let's try an example: an application of the algorithm on Ames Housing data; (3) Advantages: any advantage of the algorithm such as if it is generally easy to understand and implement or accurate; (4) Disadvantages: any drawbacks of the algorithm such as the difficulty associated with explaining its results or failures to capture the more complex relationship between the predictors and the outcome; (5) Curious to learn more: a link to the Wikipedia page of the algorithm. Participants need to click through all the algorithms before they could continue to the next page where they decided on which algorithm to use.}
\label{fig:learn_alg}
\end{figure*}

Figure ~\ref{fig:learn_alg} illustrates what the page looks like for linear regression. The explanations for the different algorithms were carefully designed by the research team using layman's terms. For each algorithm, the explanation illustrates how the algorithm works through data visualization and an example of the model trained to predict housing price using the Ames Housing data \cite{kaggle}. The description frames the advantages and disadvantages of the algorithm in terms of ease of model implementation, explainability of prediction results, model complexity, and accuracy. Lastly, we provided links to the Wikipedia page for participants who wanted a more technical description.

\begin{figure}[h]
\includegraphics[width=0.5\textwidth]{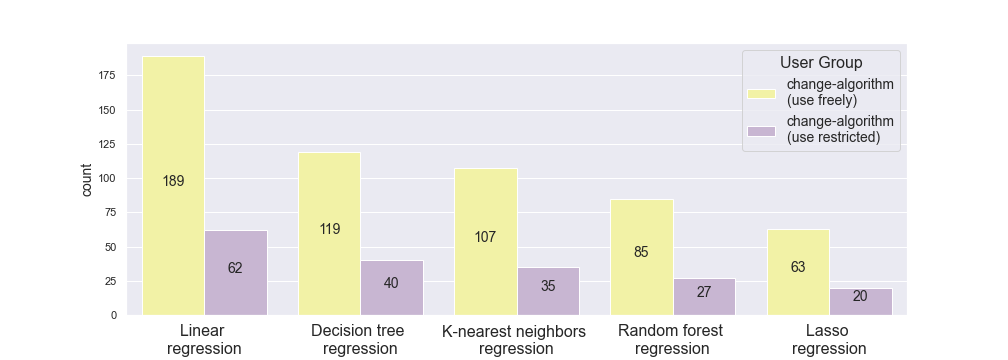}
\caption{Input Algorithms Selected by Participants. Each bar corresponds to the number of participants that chose to use the algorithm in each user group. Participants preferred linear regression the most, followed by decision trees and K-nearest neighbors.}
\Description{Input Algorithms Selected by Participants. Each bar corresponds to the number of participants that chose to use the algorithm in each user group. Participants preferred linear regression the most, followed by decision trees and K-nearest neighbors.}
\label{fig:algorithms_selected}
\end{figure}

We attempted to frame the choice of algorithm less as a technical one but more as a choice of trade-offs between the abovementioned pros and cons. This means that we do not expect participants to fully grasp the technical details of the algorithms. Rather, we expect them to form some intuition of the working of the algorithms to make decisions by weighing the pros and cons. Figure ~\ref{fig:algorithms_selected} shows that most participants prefer linear regression followed by K-nearest neighbors and then decision trees. In the text responses, participants overwhelmingly indicated that they prefer the most understandable algorithms. 

Participants were required to view explanations on all the algorithms before they would choose one on the next page.  They could go back to review the descriptions of the algorithms before they committed to their choice. Participants then waited for 15 seconds while the model was being constructed. 

The resulting models' performance does vary depending on what inputs and what algorithms the participants choose, but not by very much. Figure ~\ref{fig:simulation_vars} and Figure ~\ref{fig:simulation_algs} in Appendix ~\ref{appendix:model_performance_given_design} show simulations of the performance of possible models that can be constructed based on different choices of variables or training algorithms, respectively. Overall, these performance metrics reflect the reality that, in most circumstances, there are many models that perform nearly optimally and any gains from particular design choices are minimal.  It is of course possible to degrade performance by choosing only a small set of inputs that have low predictive power.  

\subsubsection{Bonus Schemes}
\label{section:bonus_schemes}

Next, all participants learned about how they would earn up to \$5 in bonus compensation depending on the accuracy of their predictions according to the scheme shown in Table ~\ref{table:bonus_scheme}. The payment scheme used on Prolific is available in Appendix~\ref{appendix:prolific_replication}. 
In Study 1, all participants earned a one-time participation reward of \$2. The participation reward was subsequently raised to \$4  for participants in Study 2 and 3 to reflect the longer time they spent.  

\begin{table}[h]\small
\caption{Bonus Schemes}
\begin{tabular}{cc}
\hline\hline
Bonus                      & Requirement
\\
\hline
{\$5} & within 5 points of students' actual performance on average \\
{\$4} & within 10 points of students' actual performance on average \\
{\$3} & within 15 points of students' actual performance on average \\
{\$2} & within 20 points of students' actual performance on average \\
{\$1} & within 25 points of students' actual performance on average \\
\hline\hline
\multicolumn{2}{l}{\footnotesize One-time participation fee is \$2 in study 1.}\\
\multicolumn{2}{l}{\footnotesize One-time participation fee was raised to \$4 for participants in study 2 and 3}\\
\multicolumn{2}{l}{\footnotesize to reflect their time spent.}\\
\end{tabular}
\label{table:bonus_scheme}
\end{table}

\subsubsection{Outcome Control Intervention}
All participants (except the ones in use-freely) then proceeded to chose between using the model's predictions exclusively or their own predictions exclusively for completing the task after reviewing the models' performance. Participants in use-freely condition automatically received the model's predictions and could adjust them however they liked, which resembles most use cases in real life. 

Participants in can't-design (use freely), change-input (use freely) and change-algorithm (use freely) were told that if they decide to use the models they could change the models' predictions freely.  Otherwise, they would need to rely exclusively on their own judgement. In the change-input (use freely) group, participants would view \textit{only} the variables they selected.  In the change-algorithm (use freely) group, participants could view all the variables associated with the students. For both conditions, the only difference in the information received between participants who chose to use the model or not is the availability of model predictions. 

Participants in adjust-by-10 were informed that they could adjust the model's predictions up or down by at most 10 percentiles. 

Participants in can't-change-outcome, can't-design (use restricted), change-input (use restricted) and change-algorithm (use restricted) learned that if they chose to use the models, they would be unable to change the models' predictions. 

There are two primary reasons for why we showed users the expected performance of the model (on a separate test set) prior to asking them to choose whether to use the models.  First, we want to mitigate the trade-off between procedural fairness and model accuracy \cite{Zafar_Gummadi_Weller_2018} by showing the performance to participants up front so they can choose to rely on the model only when the accuracy is acceptable to them. Secondly, it mimics the real-life concern that some problems should not be solved by algorithmic models in the first place, and acknowledges that not using algorithmic models is also a viable choice that may be informed by knowledge of model operating characteristics. 

\subsubsection{Post Survey}
Upon completion, all participants were asked to estimate their average error and confidence in their predictions. Additionally, for those who chose to use the models, they were asked to estimate the model's average error and confidence in model's predictions. All were asked to justify why they chose to use (or not use) the model. For change-input and change-algorithm groups, participants were also asked to provide a short justification for their choices of variables and training algorithms, respectively.

All participants answered survey questions on their perception of how well the models represent their assessment of the students' performance and how transparent the models' prediction processes were. They then answered questions about procedural fairness using a five-point Likert scale based on three scenarios: (1) would it be fair for the school to allocate tutoring resources to the students that the model predicts will have the lowest reading scores? (2) to recommend students with the highest predicted reading scores for a competitive scholarship in reading? (3) to decide some part of the students' final grade if the students were unable to attend exams. They would then explain their answers. The fairness questions were situated in concrete decision-making contexts and vary from low to high stakes, so we can get repeated measurements. It is important to note that the fairness question does not ask participants about their perceived fairness of a specific model component (e.g. the design interactions, resulting models, models' predictions, interactions with the models, etc.). Although understanding the distinctions would help us further unpack the mechanisms of how users formed their perceptions, the focus here is rather if participants think it is fair to use the model in the decision procedures. 

Participants then answered questions about how likely they would use the models in the future, and in an open-ended question if they had any other thoughts and feelings about the models. Lastly, they voluntarily self-reported their demographics including age, gender, race, ethnicity, highest-education level obtained, confidence in math, as well as their prior experiences in doing algorithm-related studies on online platforms. 

\section{Results}
\label{section:results}
\subsection{Replication on Outcome Control (RQ 1)}
\label{section:study1}
\subsubsection{MTurk Replication}
\label{section:dietvorst_replication}
\begin{figure}[h]
\includegraphics[scale=0.4]{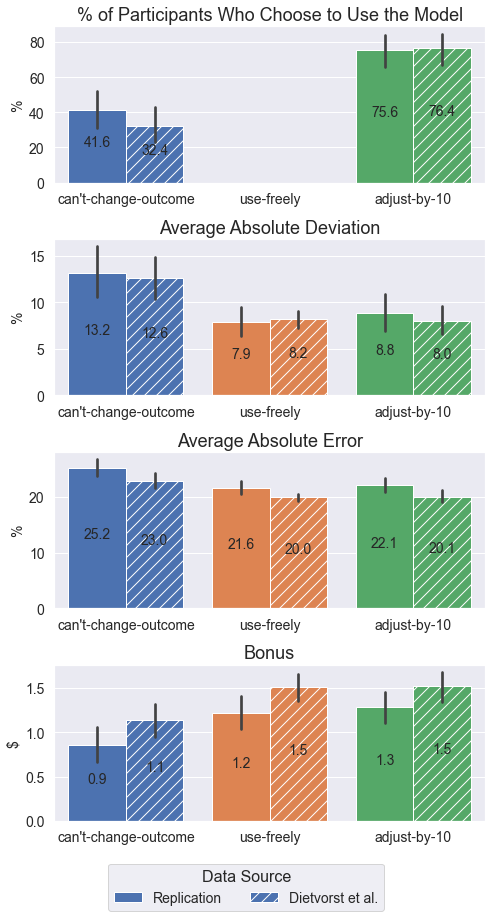}
\caption{Comparing Findings From Replication and Dietvorst et al. The non-shaded bars and shaded bars are results from our replication of Dietvorst et al.'s study and the original study respectively. We successfully replicate the findings that allowing participants to change outcomes significantly increases their likelihood of choosing the model, reduces deviation from the models' predictions, and lowers prediction errors.}
\Description{Comparing Findings From Replication and Dietvorst et al. The non-shaded bars and shaded bars are results from our replication of Dietvorst et al.'s study and the original study respectively. We successfully replicate the findings that allowing participants to change outcomes significantly increases their likelihood of choosing the model, reduces deviation from the models' predictions, and lowers prediction errors.}
\label{fig:study1_comparison_summary}
\end{figure}

For RQ\ref{rq:replication}, our Study 1 successfully replicates the findings in \citet{Dietvorst18} as shown in Figure ~\ref{fig:study1_comparison_summary}.  The figure presents our results in solid bars and results extracted directly from \cite{Dietvorst18} in adjacent shaded bars. 

We find that allowing participants to adjust the outcomes statistically significantly increased their likelihood to use the algorithms and lowered prediction errors. Only 41.6\% of the participants in the can't-change-outcome group chose to use the model compared to 75.6\% ($\chi^2(1, N=159) = 19.05, p < 0.000$) in the adjust-by-10 group. The can't-change group also on average deviated much more from the models' predictions than others. The average absolute deviation from model was 13.2\% in can't-change group compared to 7.9\% in use-freely group ($t(149)= 3.22, p = 0.002$) and 8.8\% in adjust-by-10 ($t(157)= 2.51, p = 0.013$) group.

Participants who could modify outcomes committed significantly lower errors than those who could not. The AAEs were 21.6\% for the use-freely ($t(149)=3.49, p < 0.000$) group and 22.1\% for adjust-by-10 ($t(157)=2.96, p = 0.004$) group ---both are significantly lower than the AAE of 25.2\% for the can't-change-outcome group. Subsequently, both groups earned higher bonuses.

\begin{figure}[h]
\includegraphics[width=0.5\textwidth]{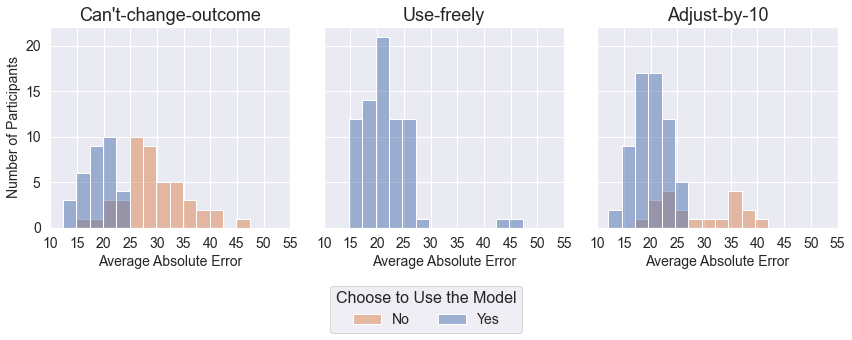}
\caption{Distribution of Average Absolute Errors by Participants' Choice to Use the Model. The blue and red bars represent the average absolute error distributions for participants who chose to use the models or not respectively. For all conditions, the error distributions for participants who chose to use the model appear left skewed, suggesting they are less likely to make large errors.}
\Description{Distribution of Average Absolute Errors by Participants' Choice to Use the Model. The blue and red bars represent the average absolute error distributions for participants who chose to use the models or not respectively. For all conditions, the error distributions for participants who chose to use the model appear left skewed, suggesting they are less likely to make large errors.}
\label{fig:study1_error_distribution_summary}
\end{figure}

Figure ~\ref{fig:study1_error_distribution_summary} shows the distribution of AAEs by whether participants chose to use the model or not. For all conditions, the error distributions for participants who chose to use the model appear left skewed, suggesting they are less likely to make large errors. 

\noindent \textbf{\textit{An important caveat.}} The overall story with replication is not so clean when one takes into consideration results from Study 2.  As mentioned in \textsection\ref{section:RQs}, the \textit{can't-change-outcome} condition of Study 1 is identical to the \textit{can't-design (use restricted)} condition of Study 2.  The latter (identical) condition run on MTurk is shown in the solid red bars in Figure~\ref{fig:all_study_summary}.  Here we find that 67.4\% of participants in the condition choose to use the model, compared to just 41.6\% in the identical condition in Study 1.  The conditions were identical except for being run on different days.  The difference is not only statistically significant, but is on par with the difference observed between the \textit{can't-change-outcome} baseline condition and the \textit{adjust-by-10} outcome control condition in Study 1 (and \cite{Dietvorst18}).  What we observe is essentially a massive and unexpected batch effect.  Time of day and serial positioning may have influenced the results, leading to contradictory findings \cite{casey2017intertemporal}.  Such observations contribute to pre-existing concerns surrounding the study of human-AI interaction on crowdworker platforms.  

\subsection{Effects of Process Control (RQ 2)}
\label{section:process_control}
\begin{figure*}[h]
\includegraphics[scale=0.4]{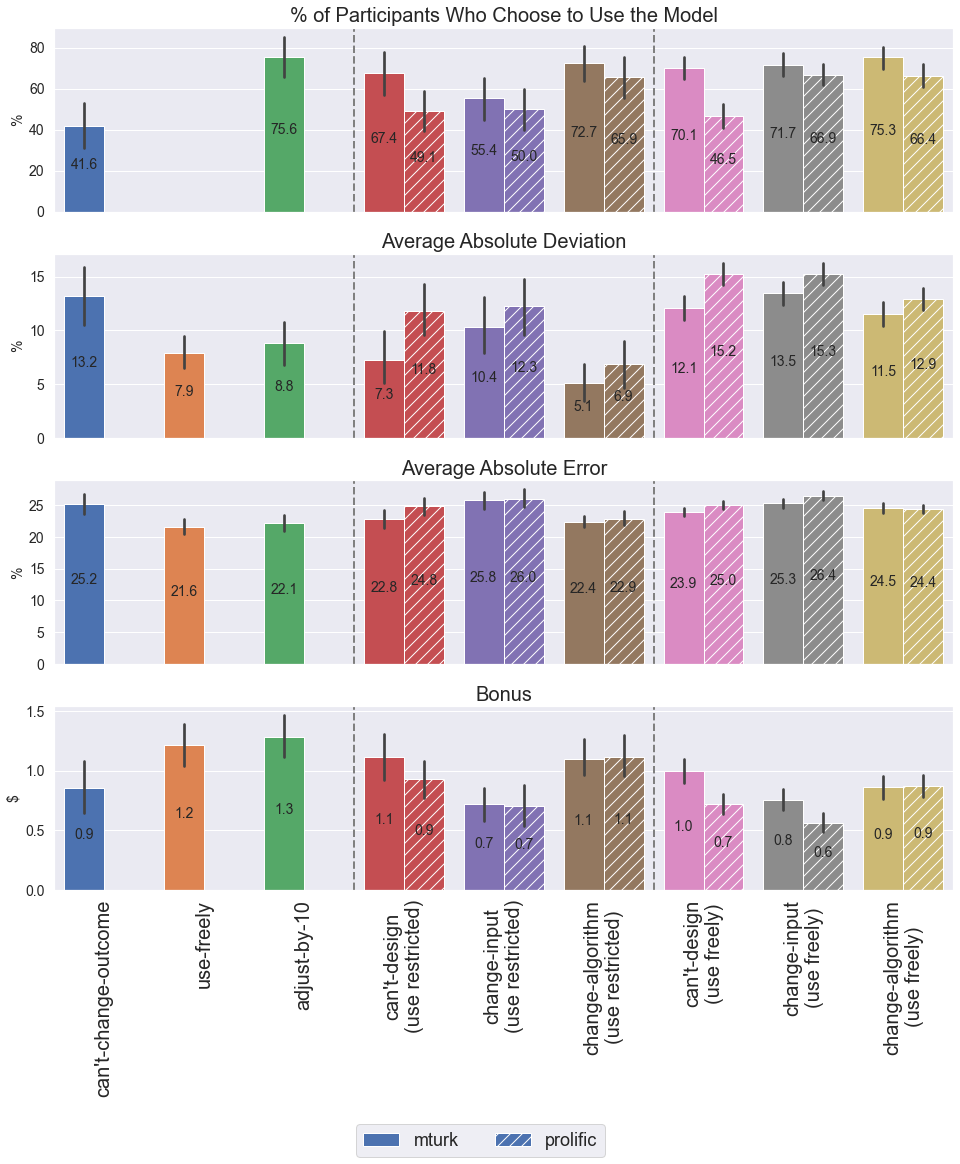}
\caption{All Experimental Conditions Results on MTurk and Prolific Platforms. The gray dotted vertical line separates each study. The left, middle, and right frames correspond to study 1, 2 and 3 respectively. Use-freely group received models without having to choose to use the models or not, which means 100\% of the participants chose to use the model by default and the corresponding bar is removed in the first row.}
\Description{All Experimental Conditions Results on MTurk and Prolific Platforms. The gray dotted vertical line separates each study. The left, middle, and right frames correspond to study 1, 2 and 3 respectively. Use-freely group received models without having to choose to use the models or not, which means 100\% of the participants chose to use the model by default and the corresponding bar is removed in the first row.}
\label{fig:all_study_summary}
\end{figure*}

In the second study, we address RQ 2: What is the effect of process control on algorithm aversion? We first conducted experiments MTurk and then ran a replication study on Prolific. Figure ~\ref{fig:all_study_summary} summarizes the results from the experimental conditions from all three of our studies, where the gray dotted vertical lines separate each study. The three sets of bars in the middle frame correspond to the Study 2 can't-design/change-input/change-algorithm (use restricted) respectively, where the non-shaded and shaded bars correspond to results on MTurk and Prolific respectively. 

As noted in our discussion of the Study 1 results, the \textit{can't-design (use restricted)} condition of Study 2 on MTurk does not replicate the identical experimental condition \textit{can't-change-outcome} of Study 1. We also see significant differences between the MTurk and Prolific Study 2 results for the \textit{can't-design (use restricted)} condition.  Whereas the model take-up rate on Prolific is 49.1\% and statistically similar to the 41.6\% observed on MTurk (Study 1), it is significantly different from the 67.4\% rate observed on MTurk in Study 2.  These batch and platform differences complicate the interpretation of the results. In reporting results for Study 2 and 3 we will primary discuss our results on MTurk, and qualify the findings with our observations from Prolific. 

\subsubsection{MTurk Results}
When we compare \textit{only within Study 2}, we find that when participants had no outcome control, providing them with process control in the form of input variable selection did not significantly affect their likelihood to choose to use the models.  Participants were more likely to choose to use the models if they could change the algorithms compared to if they could not design at all, but the observed difference is not statistically significant on MTurk. As shown in Figure ~\ref{fig:all_study_summary}, 55.4\% ($\chi^2(1, N=178) = 2.70, p = 0.1$) of the participants who could change the inputs and 72.7\% ($\chi^2(1, N=185) = 0.62, p = 0.433$) of the participants who could change the algorithms chose to use the models compared to 67.4\% in the can't-design group.

The change-input (use restricted) group deviated more from the models by 10.4\% compared to the can't-design (use restricted) group which only deviated from the model by 7.3\% ($t(176)= -1.69, p = 0.091$). The change-input (use restricted) group also committed a significantly higher AAE of 25.8\% compared to the can't-design group of 22.8\%($t(176)= -3.20, p = 0.002$). 

The change-algorithm (use restricted) group, on the other hand, appeared to be more likely to use the model, made lower errors, and deviated less from the model compared to the can't-design (use restricted) group, although the differences are not statistically significant. 

Comparing between the two process control groups, we observe large and statistically significant differences between change-input and change-algorithm conditions in their likelihood to use the model ($\chi^2(1, N=191) = 6.22, p =  0.013$), deviation ($t(189)= -3.36, p = 0.001 $), and prediction performance ($t(189)= -4.30, p < 0.000$).  Thus, being able to select the training procedure significantly increases model take-up compared to being able to select which input features will be used by a pre-determined training process.  

Subsequently, we observe that the can't-design (use restricted) group earned a higher average bonus than change-input (use restricted), although this is mostly an artifact from the bonus payment scheme---the 25.8\% AAE of the change-input group on MTurk borders the cutoff at 25\% for earning \$1. 

When we instead compare to the can't-change-outcome condition of Study 1, which was also conducted on MTurk and is identical to the can't-design (use restricted) condition of Study 2, our results remain qualitatively the same but statistically different. In this comparison, we maintain the finding that being able to change model inputs does not significantly increase take-up rates (55.4\% vs. 41.6\%, ($\chi^2(1, N=169)=3.23, p = 0.072$)). However, being able to change the training algorithm significantly mitigates algorithm aversion, bringing model take-up rates to 72.7\% ($\chi^2(1, N=176)=17.41, p < 0.000$).

\subsubsection{Replication Results from Prolific}
We then conduct the exact same study on Prolific with a few platform-specific adaptions noted in Appendix ~\ref{appendix:prolific_replication}. The three shaded bars in the middle frame in Figure ~\ref{fig:all_study_summary} correspond to the replication results of can't-design/change-input/change-algorithm (use restricted) respectively. Comparing the MTurk (non-shaded) and Prolific (shaded) results, one may immediately notice that the magnitudes of the results vary considerably across platforms. 

Replication on Prolific confirms that given no outcome control, there is no statistically significant difference between participants in the can't-design and the change-input groups in their likelihood to choose the models and that the change-algorithm group was more likely to choose the model compared to the can't-design group. Whereas the difference between can't-design (use restricted) and change-algorithm (use restricted) is not statistically significant on MTurk---mainly due to the aforementioned unexpectedly high take-up rate of 67.4\% in the baseline condition in this batch---it is significant on Prolific ($\chi^2(1, N=197)=5.53, p = 0.019$), and when assessed with respect to the identical Study 1 condition. 

The change-input (use restricted) group again deviated more from the models' predictions by 12.3\% and made a higher AAE of 26\%, compared to the can't-design (use restricted) group which deviated by 11.8\% and made an AAE of 24.8\%, although these differences are not statistically significant. 

Similar to the results on MTurk, the change-algorithm (use restricted) group deviated significantly less from the model ($t(195) = 2.85, p = 0.005$) and performed significantly better on the tasks ($t(195) = 2.03, p = 0.043$) compared to the can't-design (use restricted) group. 

The Prolific experiments confirms the finding on MTurk that there are significant differences in these primary outcomes between change-input/algorithm (use restricted) as well. This suggests that the two types of interventions have different impacts on how participants interact with the models.

\subsection{Comparing Outcome and Process Control (RQ 4 and 5)}
\label{section:comparing_outcome_process}
We now turn to the Research Questions 4 and 5: How does the effect of process control compare to the effect of outcome control when it comes to mitigating algorithm aversion?  Does providing \textit{both} process and outcome control further reduce algorithm aversion beyond the effects of each control separately?  In the previous section we saw that participants were more likely to choose the model when they could design the algorithm \textit{or} when they could deviate (even slightly) from the model's predictions, as compared to when they had no process or outcome control.  To understand how process and outcome control interact we ran three additional study conditions that are similar to those in Study 2, except that we now allowed participants to freely modify the model's predictions. As we discuss, our findings on the Prolific platform diverge from those reported on MTurk.  Results for Study 3 are once again summarized in Figure ~\ref{fig:all_study_summary}, where the last three grouped bars correspond to the Study 3 can't-design (use restricted), change-input (use restricted), and change-algorithm (use restricted) conditions, respectively.   

In Appendix~\ref{appendix:app_regression} we present the results of a regression analysis that pools data across study conditions.  Because of the issues noted in replicating results across study conditions and platforms, the regression results are somewhat challenging to interpret.  For this reason we focus our discussion in the main text on Figure ~\ref{fig:all_study_summary}, which allows us to more easily interpret the results of our experiments.  

\subsubsection{MTurk Results}
From the Study 3 results summary in Figure ~\ref{fig:all_study_summary}, we see that 71.7\% of participants who could change inputs and use model predictions freely chose to use the model.  This is statistically indistinguishable from the model take-up rate of 75.6\% for participants who had no process control but could adjust predictions by 10 points (Study 1, adjust-by-10) ($\chi^2(1, N=368)=0.49, p = 0.482$).  It is also statistically indistinguishable from the 70.1\% model take-up rate of participants who had no process control but could adjust predictions freely (Study 3, can't-design (use freely)) ($\chi^2(1, N=573)= 0.16, p = 0.685$).  For participants who could change inputs, they were significantly more likely to choose to use the model when they had outcome control (71.7\%) than when they did not (55.4\%).  Taken together, these findings indicate that control over model inputs does not mitigate algorithm aversion, but that outcome control consistently increases model take-up irrespective of whether users can change model inputs.

Among participants who could change the training algorithm, 75.3\% chose to use the model when they had outcome control (Study 3, change-algorithm (use freely)), which is statistically indistinguishable from the 72.7\% of participants could modify model predictions freely (Study 2, change-algorithm (use restricted))($\chi^2(1, N=370)= 0.62, p = 0.249$). This shows no additional effect of outcome control if the participants could change the training algorithm. 

The model take-up rates for groups that could change the training algorithm are also indistinguishable from the model take-up rate for participants who had no process control but could either adjust prediction by 10 points (75.6\%, Study 1, adjust-by-10) or freely (70.1\%, Study 3, can't-design (use freely)).  Taken together with the findings from studies 1 and 2, these results show that, while outcome control and process control in the form of control over the training procedure each mitigate algorithm aversion to the same extent (RQ 4), providing both forms of control has no further mitigating effect (RQ 5).  

To summarize, on MTurk we find that participants who are able to exercise \textit{both} outcome and process control (by changing the training algorithm) are equally likely as participants who could exercise \textit{only} process control by changing the training algorithm (Study 1) or \textit{only} outcome control (Study 2) to choose to use the model.  As noted in the discussion of Study 2, process control in the form of changing inputs did not have an effect on model take-up rates when participants did not have outcome control.  In Study 3 we also find that, among participants who had outcome control, allowing participants to change inputs also did not change model take-up rates.  When participants had outcome control or could change the training algorithm, \textit{or both} they were more likely to choose to use the model, had lower deviation from model predictions, lower prediction error, and earned higher bonuses, compared to if they have no control at all. 

\subsubsection{Replication Results from Prolific}
There is a notable and unexpected difference between the MTurk and Prolific results that changes the conclusions we draw regarding the joint effect of process and outcome control.  As we will see, this difference stems primarily from the unexpectedly low model take-up rate among Prolific participants who had outcome control but no process control.    

On Prolific, we find that just 46.5\% of participants who had outcome control but no process control (Study 3, can't-design (use freely)) chose to use the model, compared to 70.1\% on MTurk.  This result is surprising not only because of the large statistically significant difference observed across platforms  ($\chi^2(1, N=591)=33.78, p < 0.000$) but because the Prolific result does not align with the findings indicated in \cite{Dietvorst18}.  While \citet{Dietvorst18} do not explicitly consider the ``use freely'' condition, based on their experimental results the authors conclude that outcome control mitigates algorithm aversion to approximately the same extent irrespective of the the degree to which participants are allowed to deviate from the model's predictions.  It is therefore surprising that on Prolific, among participants who have no process control, we find that the model take-up rate is the same for participants who cannot change model predictions (49.1\%, Study 2, can't-design (use restricted)) as for those who can change predictions freely (46.5\%, Study 3, can't-design (use freely)).  

Because we do not observe that outcome control alone mitigates algorithm aversion among Prolific participants, our conclusions about the combined effect of outcome control and process control also differ.  For instance, we find that, when participants have outcome control, they are significantly more likely to choose to use the model when they can change model inputs (66.9\%, Study 3 change-input (use freely)) compared to when they cannot (46.5\%, Study 3, can't-design (use freely)) ($\chi^2(1, N=593)=25, p < 0.000$).  Similarly, when participants can change model inputs, they are significantly more likely to choose to use the model when provided outcome control (66.9\%, Study 3, change-input (use freely)) compared to when they have no outcome control (50.0\%, Study 2, change-input (use restricted)).  Thus we observe a significant interaction effect between process control in the form of changing inputs and outcome control in our experiments on the Prolific platform.  Whereas changing inputs alone or being able to deviate freely from model predictions on its own does not mitigate algorithm aversion, having both forms of control has a strong and statistically significant mitigating effect.  

Among participants who can exercise process control by changing the training algorithm, model take-up rates are similar among participants who also had outcome control (66.4\%, Study 3, change-algorithm (use freely)) as those who had no outcome control (65.9\%, Study 2, change-algorithm (use restricted)).  Thus in addition to finding no effect of outcome control alone on model take-up rates, we do not find evidence that the additional provision of outcome control further mitigates algorithm aversion among participants who are able to change the training algorithm.

\subsection{Secondary Outcomes (RQ 3)}
Lastly, we are also interested in how process control could affect participants' perception of the resulting models (RQ \ref{rq:secondary}). 

We find no statistically significant difference in participants' perception across all the conditions.  Our findings are presented in Figure ~\ref{fig:trf_ratings} in Appendix ~\ref{appendix:supplement_figures}. This suggests that being able to control the models' design or predictions alone is not enough to improve participants' perceptions of other dimensions of the models. This stands contrary to studies involving participatory design that show designing can improve perceived fairness. This is likely because our experimental study lacks meaningful interactions between designers and users who are experienced and have a high stake in solving the problems. 

\begin{figure}[h]
\includegraphics[scale=0.35]
{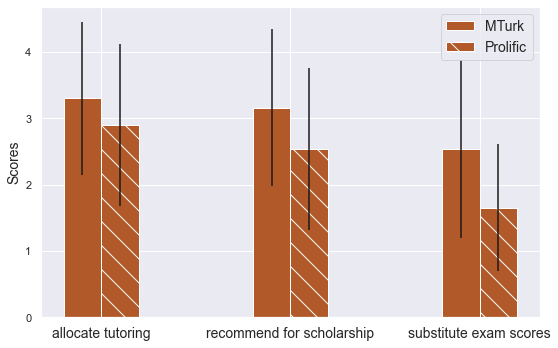}
\caption{Fairness Rating by Question. Participants on both MTurk and Prolific on average rated the model as more fair if it is used to allocate tutoring resource to students with the lowest predicted scores than if it is used to substitute exam scores.}
\Description{Fairness Rating by Question. Participants on both MTurk and Prolific on average rated the model as more fair if it is used to allocate tutoring resource to students with the lowest predicted scores than if it is used to substitute exam scores.}
\label{fig:fairness_rating_by_question}
\end{figure}

Interestingly, we observe that participants are responsive to hypothetical use cases and deem the model less fair as the decision stakes increase. As shown in Figure ~\ref{fig:fairness_rating_by_question}, participants on both platforms believed using the tool to allocate tutoring sources to students with the lowest predicted scores is on average more fair than replacing students' performance with predicted performance if they were unable to attend exams---a scenario that indeed arose in 2020 when Ofqual used an algorithm to generate students' A-level test scores when the exams were canceled during the COVID pandemic \cite{education_2020}. This finding emphasizes the importance of framing questions about the fairness of algorithmic tool as the ratings can vary when the questions are contextualized by use cases.


\section{Discussion}
In this work we studied the effect of two different kinds of process control on algorithm aversion: being able to select which inputs are used by a model, and being able to change the training algorithm used (and thereby control the structure of the resulting model).  Our work contrasts with studies on process \textit{transparency} because in our setting we not only help participants understand how the models work, but also provide them with the ability to control how the models are constructed.  As critics have argued, transparency in the sense of being able to see how a system works, is alone not sufficient for understanding and governing algorithmic systems \cite{Ananny18}. Most notably transparency can be disconnected from power to change the systems. 

We first replicated \citet{Dietvorst18}'s study on outcome control, confirming that allowing participants to change outcomes significantly increase their likelihood of choosing the model, reduces deviation from the models' predictions, and lowers prediction errors. However, we also discuss an important caveat that does not fully agree with \cite{Dietvorst18}'s results in \textsection\ref{section:study1}.  

We then show that allowing people to customize the models---even if slightly---can increase their willingness to use models. If participants are able to choose which training algorithm to use, they are more likely to use the models' predictions, deviate less from those predictions, and commit lower errors, compared to if they could not design and could not change models' predictions. However, being able to change only the models' input factors has little effect on on these outcomes. 

More importantly, we find that process control is as effective as outcome control in mitigating algorithm aversion. Allowing participants to choose which training algorithms to use can achieve similar results as allowing them to change the models' predictions by a limited amount in reducing aversion and prediction errors. 

We also find that the combination of process \textit{and} outcome control is equally effective as process or outcome control alone in mitigating algorithm aversion.  That is, the effects do not stack.  As described in the previous section, the interpretation of our results is complicated by significant cross-study and cross-platform variation.  

These findings have many important implications for workplaces that seek to increase buy-in when implementing algorithmic systems. Involving end-users in the model design process can increase their utilization of algorithms. However, system designers need to consider which design levers are likely to impact buy-in in the given setting.

Although it seems natural to gather feedback from users on what evidence should be used for the algorithm to arrive at its decisions (i.e., what the input features should be), our findings indicate that this form of process control may not mitigate algorithm aversion. This finding is surprising when considering previous work on involving users in the feature engineering process.  In particular, stakeholders and intended end-users are routinely asked to weigh in on input features in domains such as health care \cite{wang2018learning}.    The primary goals of feature engineering are to improve model performance \cite{https://doi.org/10.48550/arxiv.2103.04044, heuer2022comparative} and increase transparency and user trust \cite{10.1093/jcmc/zmac010}. Our study does not address the first goal but is related to the latter. We find no effects of changing inputs on perceived transparency, fairness, and confidence in the models. This may be potentially due to the limited engagement between designers and users, and the lack of iterative interactions that typically occurs in feature engineering, which partially explains why changing inputs does not reduce aversion. Another potential explanation we propose is based on research on deliberative public engagement \cite{pytlikzillig2018deliberative}. \citet{pytlikzillig2018deliberative} suggest that critical thinking prompts can increase the perceived quality of information received by the participants, but also increases their likelihood of rejecting unfavorable decisions. These processes can operate simultaneously, canceling each other out and producing what appears to be "null effects" on the willingness to accept a decision. As discussed briefly in \textsection\ref{section:process_control_interventions} and Figure~\ref{fig:vars_selected}, participants in the change-input groups intentionally avoid using certain variables suggesting they may have strong mental models or preferences for the resulting model. It is possible that they ultimately deviate from or refuse to use the models because the models did not meet their presumed expectations or preferences. However, further study is required to confirm this.

Based on our investigation of user selection of the training algorithm, we also suggest that system designers should not underestimate users' appreciation for learning  about the inner workings of the algorithms, even though they are not technical experts in the area. This could be an integral part of the onboarding process for users. 

It is important to note that there is a risk of using design choices that overcome aversion to intentionally obfuscate or manipulate user choice (i.e., ``dark patterns''). Nudging users to trust a harmful predictive system is an undesirable practice. This is especially dangerous when an illusion of being able to modify the system can also create a sense of control \cite{Vaccaro18}. Additionally, users can commit commission errors when they follow the algorithmic recommendations without taking into account other sources of information. Many real-life cases of algorithm-assisted decision-making demonstrate the critical role of users in interrogating, investigating, and critically incorporating algorithmic advise. When algorithmic system failure occurs, experienced workers have been shown to be capable of identifying and correcting for inaccurate predictions to achieve better performance \cite{de2020case}. Human-in-the-loop decision-making has also been found to reduce disparities compared to fully automated hypotheticals \cite{cheng2022child}. Although we focus on reducing algorithm aversion in the study, we caution the readers that successful human-AI collaboration requires proper- instead of over-reliance on algorithms. 

Lastly, although we found some consistent results across MTurk and Prolific, we also observed statistical significantly differences across platforms and within the same platforms across time. Although we provided a rather limited range of changes participants could make to ensure the intentions of our studies are faithfully communicated across platforms, participants can form different expectations of the models absent engagement with the researchers and designers, leading the results to be dependent on their observed or unobserved characteristics. As mentioned, the time of day and serial position are associated with small but important variations in demographic composition \cite{casey2017intertemporal}. There is also evidence that a non-trivial percentage of Turkers falsify demographic and personality traits in order to participate in research from which they would otherwise be excluded \cite{fowler2022frustration}.  As we report on in the regression modeling results presented in Appendix~\ref{appendix:app_regression}, controlling for demographic characteristics does not help bridge cross-platform and cross-time differences identified observed our experiments.   Given the growing concerns regarding replicability and reproducibility in HCI in recent years \cite{wilson2013replichi, 10.1145/3170427.3188395, 9585852}, our study follows the recommended best practices by employing the ``replication and extension'' model and serves as a cautionary tale in soliciting feedback from crowdworkers.

\section{Limitations}
Given that our studies were conducted on crowdworkers from MTurk and Prolific, it is necessary that we qualify the interpretation of the results. Ideally, the participants should embody end-users who currently do not have the authority or capacity to make decisions about how an algorithmic system should be designed, but have domain expertise and will use the end product every day in their decision-making. Crowdworkers, however, are very different from real algorithm users, notably in their lack of relevant domain expertise and non-monetary incentives. In practice, experience has been found to influence how users interact with algorithmic systems and the ability to engage in appropriate reliance \cite{cheng2022heterogeneity, de2020case,cheng2022child}. 

Furthermore, in practice stakeholders are able to provide input into model design through interactive methods such as interviews, workshops, or co-design sessions.  Such forms of engagement may compound with various forms of process control in ways that we were unable to investigate in our study.

\section{Future Work}
While the choice of input features and model structure are key elements of AI system design, prior work on problem formulation has shown how the choice of target variable is often an important, subjective, and value-laden decision \cite{passi2019problem}.  For instance \citet{Obermeyer447} showed how predicting health care costs rather than direct measures of health care needs results in tools that may disadvantage Black patients.  It is thus of interest in future work explore how allowing participants to select the outcome measure (e.g., by specifying what target variable to predict, or how to combine competing risk measures into a single index) influences algorithm aversion. 

Future work could also further test the underlying psychological mechanism that explains the observed effects. There could be many potential and plausible theories such as "IKEA effect", mere exposure effect, illusory truth, effort justification, and psychological ownership, to name a few. Identifying most effective underlying mechanisms would allow practitioners to better understand which forms of process control are likely to mitigate algorithm aversion in which settings.

Lastly, given the reproducibility challenges highlighted throughout our study, we welcome further replication of our work.  We facilitate re-analysis of our data and further replication of our experiments by providing the data, analysis code, and links to the survey platform in the supplement.

\begin{acks}
This material is based upon work supported by the National Science Foundation under Grant 1939606. Any opinions, findings, and conclusions or recommendations expressed in this material are those of the author(s) and do not necessarily reflect the views of the National Science Foundation. We would like to thank Sreyantha Chary Mora for creating the survey platform and making the study possible. We also want to thank the many referees of the previous versions of
this paper for their extremely useful suggestions.
\end{acks}
 
\bibliographystyle{ACM-Reference-Format}
\bibliography{algorithm_aversion_process_control_lit}

\clearpage
\appendix
\section{Recruitment}
\label{appendix:recruitment}
For study 1, we aimed to recruit for 400 participants with 100 participants for each condition. The final sample size we obtained after applying the exclusion criteria is 321, which is comparable to \cite{Dietvorst18}. We aimed to recruit 100 participants for each condition in study 2 as well. In study 3, we recruited for 300 participants per condition so we can increase our sample size and better detect any interaction effects. The numbers of participants for all study conditions by platforms are summarized in Table~\ref{table:recruitment}.

\begin{table*}[h]
\caption{Number of Participants by Platform and Study Conditions}\small
\begin{tabular}{@{}cccccccc@{}}
\toprule
                          & Group Name                             & \multicolumn{3}{c}{\# Participants}                                               & \multicolumn{3}{c}{\# Participants (Passed Attention Checks)}                      \\ \midrule
study                     &                                        & Mturk                     & Prolific                  & Pooled                    & Mturk                     & Prolific                  & Pooled                    \\
\multirow{4}{*}{1}        & can't-change-outcome                   & 77                        &                           & 77                        & 59                        &                           & 59                        \\
                          & use-freely                             & 74                        &                           & 74                        & 53                        &                           & 53                        \\
                          & adjust-by-10                           & 82                        &                           & 82                        & 60                        &                           & 60                        \\
                          & adjust-by-10 (proposed bonus   scheme) & 88                        &                           & 88                        & 73                        &                           & 73                        \\
\hline
\multirow{3}{*}{2}        & can't-design (use restricted)          & 86                        & 112                       & 198                       & 58                        & 103                       & 161                       \\
                          & change-input (use restricted)          & 92                        & 90                        & 182                       & 58                        & 86                        & 144                       \\
                          & change-algorithm (use   restricted)    & 99                        & 85                        & 184                       & 64                        & 73                        & 137                       \\
\hline
\multirow{3}{*}{3}        & can't-design (use freely)              & 288                       & 303                       & 591                       & 199                       & 281                       & 480                       \\
                          & change-input (use freely)              & 286                       & 290                       & 576                       & 199                       & 270                       & 469                       \\
                          & change-algorithm (use freely)          & 271                       & 292                       & 563                       & 190                       & 265                       & 455                       \\
                          \hline
Total              &                                        & 1,443     & 1,172       & 2,615     & 1,013             & 1,078                & 2,091             \\ \bottomrule
\end{tabular}
\label{table:recruitment}
\end{table*}

\section{Model Performances Given Design Choices}
\label{appendix:model_performance_given_design}
\begin{figure}[h]
  \begin{minipage}[c]{0.4\textwidth}
  \centering
    \includegraphics[width=\textwidth]{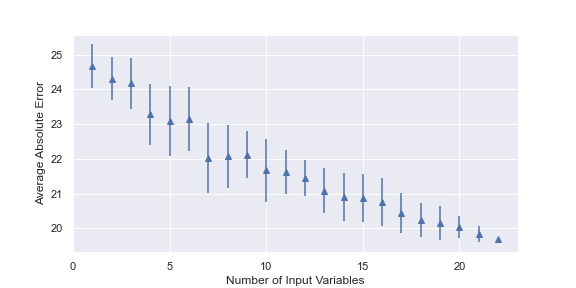}
    \subcaption{Simulated AAE Given Number of Variables Chosen. Model performance increases and standard error decreases as the number of variables chosen increases.}
     \label{fig:simulation_vars}
  \end{minipage}
  \begin{minipage}[c]{0.4\textwidth}
  \centering
   \includegraphics[width=\textwidth]{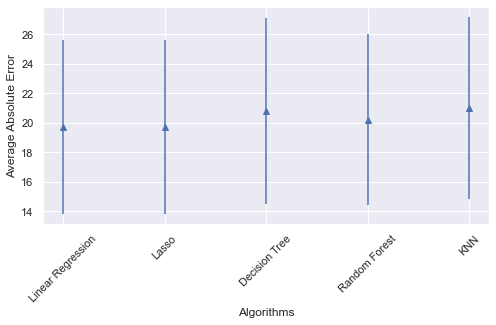}
    \subcaption{Simulated AAE Given Algorithms Chosen}
    \label{fig:simulation_algs}
  \end{minipage}
  \caption{Simulated Performance for Choosing Number of Variables and Algorithms. The error bars are the standard errors of the algorithms. (a) On average the performance ranges from 24.46\% average absolute error to 19.7 for choosing input variables. Choosing more variables leads to better performance but it is not always true. (b) Choosing algorithms result in similar performance.}
\Description{Simulated Performance for Choosing Number of Variables and Algorithms. The error bars are the standard errors of the algorithms. (a) On average the performance ranges from 24.46\% average absolute error to 19.7 for choosing input variables. Choosing more variables leads to better performance but it is not always true. (b) Choosing algorithms result in similar performance.}
\end{figure}

We simulated the performance of choosing input variables by randomly selecting N variables, training the linear regression, and recording the test average absolute errors. We repeat the exercise 20 times for N from 1 to 22 to obtain ranges for the model's performance. The performance is presented in Figure ~\ref{fig:simulation_vars}. The more variables a participants uses the better the performance of the algorithm, but it is not always true. To obtain simulated performance of choosing algorithms, we randomly select 20 samples for reach algorithm and repeat this 30 times to obtain mean average absolute error and corresponding standard errors. The result is in ~\ref{fig:simulation_algs}. In either case, on average the performance ranges from 24.46\% average absolute error to 19.7\%.

\section{Adjust-by-10 (Alternative Bonus Scheme)}
\label{appendix:adjust_by_10_alt}
When planning our replication study we noticed that in \cite{Dietvorst18}, while the maximum bonus was stated as \$5, it was statistically impossible to obtain a bonus greater than \$3. Specifically, participants were told that could earn \$5 if their predictions are within 5 points of students' actual performance on average, which would be difficult to do given the model's average absolute error rate was 17.5 points.  Given that models generally outperform humans on prediction tasks it is very unlikely that a participant could achieve an error under 5 points as needed to obtain the \$5 bonus. Under loss aversion bias \cite{kahneman1991anomalies}, participants may prefer avoiding losses by choosing to rely on a model that already (nearly) ensures a \$3 bonus rather than risk ``losing'' this bonus by making their own predictions.  By contrast, if the model's stated performance is sufficient only for a minimal \$1 bonus, they may feel there is little to lose (and up to \$4 to gain) by making predictions themselves.  To test this hypothesis we propose a new payment scheme that lowers the one-time participation reward and centers the \$3 average bonus on the models' average performance.  The proposed scheme is constructed to yield approximately the same expected total reward as the original, but offers greater incentive to choose the model because not doing so risks ``losing'' the \$3 bonus achievable by the model. Table~\ref{table:proposed_bonus_scheme} illustrates the original and proposed bonus schemes.

\begin{table*}[h]\small
\caption{Bonus Schemes}
\begin{tabular}{ccc}
\hline\hline
Bonus                      & Original Scheme (Dietvorst et al)                             & Proposed Scheme 
\\
\hline
{\$5} & within 5 points & within 14 points of students' actual performance on average \\
{\$4} & within 10 points & within 17 points of students' actual performance on average \\
{\$3} & within 15 points & within 20 points of students' actual performance on average \\
{\$2} & within 20 points  & within 23 points of students' actual performance on average \\
{\$1} & within 25 points   & within 26 points of students' actual performance on average \\
\hline\hline
\multicolumn{3}{l}{\footnotesize One-time participation fee is \$2 under original scheme and \$1 under proposed scheme during study 1}\\
\end{tabular}
\label{table:proposed_bonus_scheme}
\end{table*}

\begin{figure}[h]
\includegraphics[scale=0.3]{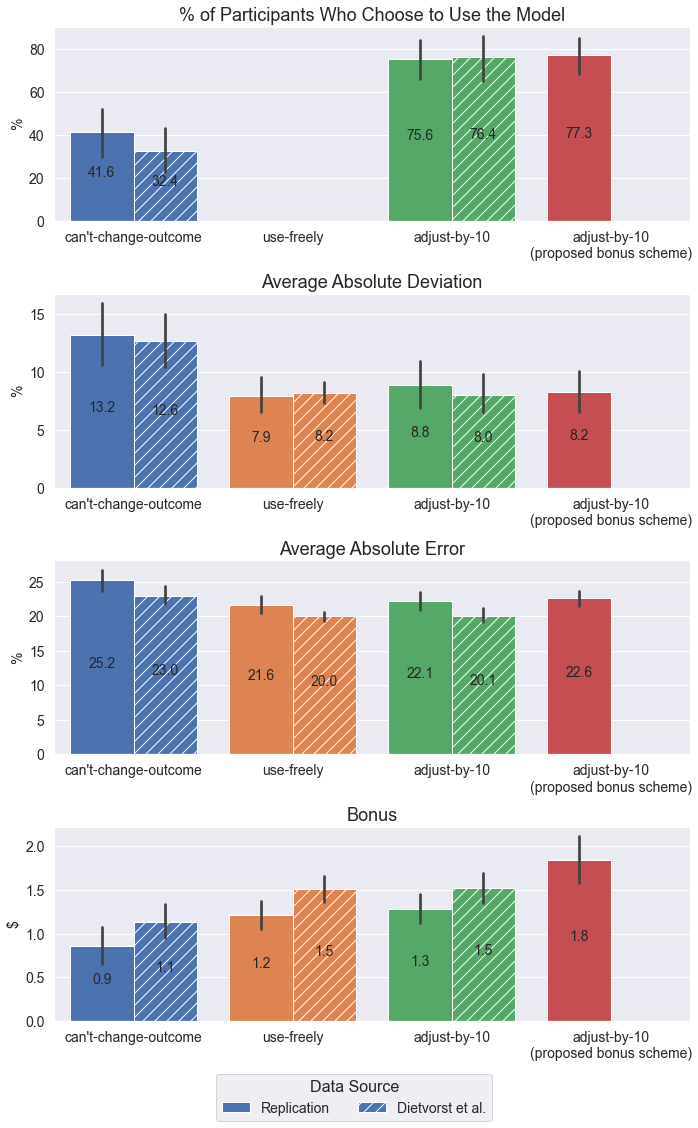}
\caption{Comparing Findings From Replication and Dietvorst et al. The non-shaded bars and shaded bars are results from our replication and Dietvorst et al.'s study respectively. We show that in addition to successfully replicate Dietvorst et al.'s findings, there is no statistically significantly differences among the adjust-by-10 groups under the original or the proposed bonus rules.}
\Description{Comparing Findings From Replication and Dietvorst et al. The non-shaded bars and shaded bars are results from our replication and Dietvorst et al.'s study respectively. We show that in addition to successfully replicate Dietvorst et al.'s findings, there is no statistically significantly differences among the adjust-by-10 groups under the original or the proposed bonus rules.}
\label{fig:study1_bonus_fig}
\end{figure}

The results are summarized in Fig~\ref{fig:study1_bonus_fig} along with other conditions in Study 1. We find that participants are not sensitive to incentive structures. They are just as likely to use the model if the model's stated performance is sufficient to achieve a bonus payout in the middle-to-top of the bonus range vs. if it is only sufficient to achieve a minimum bonus payout, a result that is surprising when viewed in the context of loss aversion.

\begin{figure}[h]
\includegraphics[scale=0.3]{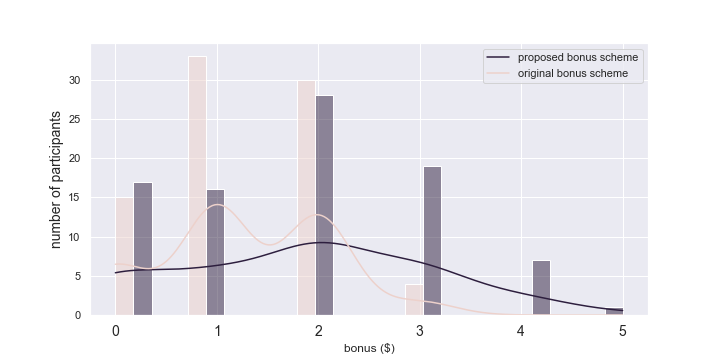}
\centering
\caption{Distributions of Bonus Earned Under Original and Proposed Bonus Schemes. Whereas no participants were able to earn more than \$3 under the original bonus scheme, participants were able to earn \$3 and above under the new scheme.}
\Description{Distributions of Bonus Earned Under Original and Proposed Bonus Schemes. Whereas no participants were able to earn more than \$3 under the original bonus scheme, participants were able to earn \$3 and above under the new scheme.}
\label{fig:study1_bonus_distribution_comparison}
\end{figure}

By construction, more participants were able to earn higher bonuses using the proposed payment as shown in Figure ~\ref{fig:study1_bonus_distribution_comparison}. Whereas no participants were able to earn more than \$3 under the original bonus scheme, participants were able to earn \$3 and above under the new scheme. Note that although participants earned on average \$0.56 more in bonus under the proposed scheme, they made \$0.44 less in the total compensation due to the lower base payment.

Based on the text responses to why they chose to use the models, only a few respondents who were subjected to the proposed bonus scheme explicitly mentioned that they chose the models because of the bonus scheme. For example, one participant said that "based on the average of being off by 19 points it meant I would get a bonus that was in the middle so that would be fine." This indicates that the change in bonus scheme did not go without notice. However, the majority of the participants were not aware of the implication of the proposed bonus scheme. 

In addition to the debatable replicability of loss aversion indicated by recent studies \cite{yechiam2019acceptable, sanders2021loss}, our null finding in this case highlights the challenges of conducting such studies online and in the unique context of human-AI collaboration. Interestingly, we note that a previous study on risk assessment instrument in judicial decisions did not identify anchor effect, another classical psychology finding, via crowdsourcing either \cite{10.1145/3479572}. Because real-life end-users can have very different demographics characteristics and non-monetary incentives and operate in higher-stake environments, we cannot reliably generalize the finding to real workplaces but acknowledge that our finding has implications on designing incentives for crowdworker studies.

\section{Prolific Replication}
\label{appendix:prolific_replication}
We used slightly different recruitment criteria and payment schemes on Prolific. 

Recruitment: Since Prolific existed for a far shorter period of time than MTurk, the same exclusionary criteria would result in too few eligible workers. Thus, we kept the first two criteria requiring participants to (1) live in the US (2) have a submission approval rate larger or equal than 97\%, but only required them to (3) have competed at least 60 HITs. Participants who have participated repeatedly in the same study or in a previous study were excluded.

Payment: Prolific suggests a higher baseline payment for all its workers. All participants received a flat rate of \$4 for participation. Additionally they had the opportunity to earn \$1 - \$5 according to the payment scheme mentioned in \textsection\ref{section:bonus_schemes}.

\section{Regression analysis}
\label{appendix:app_regression}
In this section we revisit our main research questions by presenting the results of a regression analysis that pools data across studies.  This allows us to more directly decompose the main and interaction effects between outcome and process controls. Additionally, it allows us to take into account the model errors shown or experienced by the users. While changing inputs or the training algorithm will not necessarily improve the default model's performance, there is some trade-off between model accuracy and customization. Because participants were informed of the models' performance before they were prompted to choose whether to use the models, we expect them to only use the models when the stated performance level was acceptable. Thus stated model performance is a potentially significant confounder in assessing the effects of process control. Lastly, we control for demographic variables including age, race, ethnicity, prior experience with algorithm studies, pronoun (gender), highest level of education, and confidence in math. To control for such factors and pool data appropriately we apply linear probability regression models and answer our RQ \ref{rq:design}, \ref{rq:comparison} and \ref{rq:interaction}. The notation used for the regression variables are summarized in Table ~\ref{table:regression_notation}.

There are several important caveats to the interpretation of the results. First, the studies were not conducted simultaneously and were affected by significant batch effect discussed in the main texts. By pooling the data from across the studies for each platform, we erase the temporal aspect of it. Secondly, because there are two identical conditions in two separate studies, we essentially combine them into one large treatment group in the regression analysis. This means that for example, the average likelihood of choosing to use the model for a condition where users cannot design and cannot change predictions is the average of "can't-change-outcome" (41.6\%) and "can't design (use restricted) (67.4\%). As noted in the discussion in the main texts, the outcomes of the two groups were not replicated across time. By averaging them, we create a control group with larger sample size and moderated batch effect, affecting the comparison to other groups who are only subject to a particular batch effect. For these reasons, we will refer to the magnitudes of the coefficients but remind the readers that they cannot be interpreted as the size of the effects -- in fact they are different from the effect size summarized in Figure ~\ref{fig:all_study_summary}. Some results may be more significant because of the averaging of conditions. 

Similarly, as we demonstrated in Appendix~\ref{appendix:adjust_by_10_alt}, there is no statistically significant differences between the two adjust-by-10 groups under the two bonus schemes listed in Table~\ref{table:proposed_bonus_scheme}. Because of the null effect of the proposed bonus scheme, we include the adjust-by-10 (proposed bonus scheme) participants in our analysis to increase sample size. 

Lastly, although the pooled sample resembles a population that is slightly more closely to the general population, it remains less diverse and more highly educated than the general population. We include the regression results on the pooled sample because of its larger sample size. However, it is difficult to interpret because the pooled sample erases the temporal and cross-platform differences.

\begin{table*}[]\small
\caption{Regression Notation}
\begin{tabular}{llp{8cm}}
\hline\hline
Variable                     & Notation        & Definition           \\
\hline
\textbf{Independent Variable} &            & \\
stated model error & (0,100) & model's average absolute errors over all students \\
observed model error & (0,100) & model's average absolute errors over the subset of 20 students the participants rated\\
change by 10         & $\in$ \{0,1\}   & 1 if participants can change the outcome by up to 10 percentile, 0 otherwise                                                                                                                           \\
change freely                & $\in$ \{0,1\}   & 1 if participants can change the outcome freely, 0 otherwise                                                                                                                                           \\
change algorithm             & $\in$ \{0,1\}   & 1 if participants can change the algorithm used by the model, 0 otherwise                                                                                                                              \\
change input                 & $\in$ \{0,1\}   & 1 if participants can change the input to the model, 0 otherwise                                                                      \\
age & $\geq 18$ & Age of the participants  \\
prior experience with algorithm studies & $\in$ \{0,1\}  &   1 if participants had previously done studies related to algorithms on the platform, 0 otherwise   \\
pronoun &  $\in$ \{0,1\}  & He/She/They/None of the above/Prefer not to answer   \\
race & $\in$ \{0,1\} & seven categories/None of above/Prefer not to answer \\
ethnicity & $\in$ \{0,1\} & Hispanic/Not hispanic/Prefer not to answer \\
highest level of eduction & $\in$ \{0,1\} & eight categories/Prefer not to answer \\
confidence in math & $\in$ \{1...5\} & 5-pt Likert scale \\
\midrule
\textbf{Dependent Variable} &                                                                 \\
use model                    & $\in$ \{0,1\}   & 1 if participants chose to use the model, 0 otherwise                                                    \\
Average Absolute Error (AAE) & (0,100) & $\frac{1}{20}\sum_{i=1}^{20} |\text{participants' predictions - students' true performance}$|
                \\
Average Absolute Deviation (AAD) & (0,100) & $\frac{1}{20}\sum_{i=1}^{20} |\text{participants' predictions - model's predictions}$| \\
\hline\hline
\multicolumn{3}{l}{\footnotesize The categorical variables including pronoun, race, ethnicity, highest level of education are one-hot encoded}\\
\end{tabular}
\label{table:regression_notation}
\end{table*}

\subsubsection{Likelihood of Choosing the Model}

We use the following specifications to identify the effects of our treatment variables, where $\beta_1$ to $\beta_7$ are coefficients of interests and $\beta_8$ is a vector of coefficients for all demographic variables:

{\begin{align} \label{eq:1}  
& \text{use model}_{i} = \beta_0 + \beta_1\cdot\text{model\_error\_rate}_i \\ \nonumber & +  \beta_2\cdot\text{change\_by\_10}_i +
\beta_3\cdot\text{change\_freely}_{i} \\ \nonumber & +  \beta_4\cdot\text{change\_input}_{i} +
\beta_5\cdot\text{change\_algorithm}_{i}  \\ \nonumber & +
\beta_6\cdot\text{change\_input * change\_freely}_{i} \\ \nonumber & +
\beta_7\cdot\text{change\_algorithm * change\_freely}_{i} \\ \nonumber & + 
\beta_8\cdot\text{demographics} + \epsilon_{i}
\end{align}}

We exclude use-freely group because the likelihood of using the models for these users is by definition 1. We use the same specification to analyze replication data from Prolific. Table ~\ref{table:outcome_likelihood_of_choosing_model_unfiltered} shows the regression results on MTurk (column 1), Prolific (column 3), and a Pooled results from both platforms (column 5) without controlling for the demographic characteristics. Column 2, 4, 6 show the results from running the same regression with additional control for demographic characteristics listed in Table~\ref{table:regression_notation}. Since we did not replicate the change-outcome-by-10 condition on Prolific, the coefficient $\beta_2$ was dropped in column 2 and 3. 

\begin{table*}[htbp]\centering
\def\sym#1{\ifmmode^{#1}\else\(^{#1}\)\fi}
\caption{Intervention Effects on Likelihood of Choosing the Model}\small
\begin{tabular}{l*{6}{c}}
\hline\hline
            &\multicolumn{1}{c}{(1)}&\multicolumn{1}{c}{(2)}&\multicolumn{1}{c}{(3)}&\multicolumn{1}{c}{(4)}&\multicolumn{1}{c}{(5)}&\multicolumn{1}{c}{(6)}\\
            &\multicolumn{1}{c}{MTurk}&\multicolumn{1}{c}{MTurk}&\multicolumn{1}{c}{Prolific}&\multicolumn{1}{c}{Prolific}&\multicolumn{1}{c}{Pooled}&\multicolumn{1}{c}{Pooled}\\
\hline
stated model error&       0.014         &       0.006         &      -0.028         &      -0.035         &      -0.008         &      -0.012         \\
            &     (0.017)         &     (0.017)         &     (0.019)         &     (0.019)         &     (0.013)         &     (0.013)         \\
[1em]
change outcome by 10 &       0.213\sym{***}&       0.203\sym{***}&                     &                     &       0.237\sym{***}&       0.231\sym{***}\\
            &     (0.051)         &     (0.052)         &                     &                     &     (0.044)         &     (0.045)         \\
[1em]
change outcome freely&       0.149\sym{**} &       0.160\sym{***}&      -0.026         &      -0.012         &       0.053         &       0.055         \\
            &     (0.048)         &     (0.048)         &     (0.055)         &     (0.055)         &     (0.036)         &     (0.036)         \\
[1em]
change input&      -0.016         &      -0.006         &       0.052         &       0.079         &       0.012         &       0.010         \\
            &     (0.069)         &     (0.070)         &     (0.077)         &     (0.078)         &     (0.051)         &     (0.051)         \\
[1em]
change algorithm &       0.169\sym{**} &       0.172\sym{**} &       0.188\sym{**} &       0.193\sym{**} &       0.173\sym{***}&       0.173\sym{***}\\
            &     (0.060)         &     (0.060)         &     (0.071)         &     (0.071)         &     (0.046)         &     (0.046)         \\
[1em]
change outcome and input &       0.011         &      -0.010         &       0.193\sym{*}  &       0.175\sym{*}  &       0.113\sym{*}  &       0.119\sym{*}  \\
            &     (0.075)         &     (0.076)         &     (0.082)         &     (0.082)         &     (0.055)         &     (0.055)         \\
[1em]
change outcome and algorithm &      -0.125         &      -0.133         &       0.028         &       0.018         &      -0.042         &      -0.041         \\
            &     (0.070)         &     (0.070)         &     (0.080)         &     (0.080)         &     (0.053)         &     (0.053)         \\
[1em]
demographics &               &     \cmark  &               &     \cmark         &              &      \cmark        \\
[1em]
\_cons      &       0.268         &       0.729         &       1.035\sym{**} &       1.305\sym{**} &       0.688\sym{**} &       0.894\sym{**} \\
            &     (0.337)         &     (0.389)         &     (0.372)         &     (0.428)         &     (0.250)         &     (0.289)         \\
\hline
\(N\)       &        1,369         &        1,369         &        1,172         &        1,172         &        2,541         &        2,541         \\
\(R^{2}\)   &       0.025         &       0.061         &       0.038         &       0.070         &       0.026         &       0.043         \\
adj. \(R^{2}\)&       0.020         &       0.039         &       0.033         &       0.045         &       0.024         &       0.030         \\
\hline\hline
\multicolumn{7}{l}{\footnotesize Prolific specification excludes change outcome by 10 because this condition was not replicated on Prolific}\\
\multicolumn{7}{l}{\footnotesize The sample includes participants who are from adjust-by-10 (proposed bonus scheme)}\\
\multicolumn{7}{l}{\footnotesize The sample excludes participants in use-freely group as their outcome is by default 1}\\
\multicolumn{7}{l}{\footnotesize Standard errors in parentheses}\\
\multicolumn{7}{l}{\footnotesize \sym{*} \(p<0.05\), \sym{**} \(p<0.01\), \sym{***} \(p<0.001\)}\\
\end{tabular}
\label{table:outcome_likelihood_of_choosing_model_unfiltered}
\end{table*}

We observe that the model's error rate does not statistically significantly affect users' preference of choosing the model on both MTurk and Prolific. This may be due to the fact that the models are only marginally better or worse than each other by a few percentage points. 

According to the results from MTurk, we find that allowing participants to change outcomes whether freely or by up to 10 percentiles statistically significantly increases their likelihood of choosing the model compared to if they have no control over either the process or the outcome. While in the data we observe that allowing participants to change the outcome by at most 10 points has a larger effect on take-up than doing so without any restriction, this difference is not statistically significant. 


As for process control, the results confirm our previous finding in \textsection\ref{section:process_control} that the effects differ depending on the design choices. On MTurk, if participants cannot change the predictions, allowing them to change input does not significantly affect their model take-up at all. If they can decide what algorithm to use, they are statistically significantly 16.9\% more likely to use the model. However, the effects of changing outcome and algorithm are only statistically significant because we created a control group whose likelihood of choosing the model is moderated. 

Consistent with results from MTurk and in main texts, Prolific results in columns (3) show that changing input alone plays an insignificant role in choosing to use the model. Changing algorithm leads to an 18.8\% increase in the likelihood of choosing the model, which is similar to that of MTurk.

Lastly, we observe different interaction effects between changing both the outcome and input vs. changing both the outcome and the algorithm of the models on the same platform and across the platforms, suggesting these interaction effects are unstable. On MTurk, changing both outcome and input has little effect on using the model. However, if participants can change both the outcome and the algorithm of the model, they appear to be 12.5\% less likely to use the model. This is the additional difference between participants who can and cannot freely change the outcome, had they were allowed to change the design of the algorithm. To make it concrete, for participants who can change both the outcome and the algorithm, their likelihood of choosing the model is 14.9\% + 16.9\% - 12.5\% = 19.3\% higher than for participants who cannot change the outcome or design at all. However, the effect of changing both the outcome and algorithm, then almost cancels out the effect of either changing outcome (14.92\%) or changing algorithm alone (16.93\%), which explains the finding in our main texts that being able to exercising \textit{both} outcome and process control by changing the algorithms has the same effect as doing either on model take-up.

On Prolific, we do not observe a significant effect of changing the outcome alone on mitigating algorithm aversion. We find that changing both the outcome and the input statistically increases the likelihood of choosing the model by 19.3\%. Changing both the outcome and the algorithm does not affect the likelihood. These confirm our findings in the main texts. 

In summary, we find consistently across platforms that model error rate has no significant effect on choosing to use the model. Changing input has little effect while changing algorithm has a positive effect on the likelihood of choosing the model. However, the effects of interactions between design choices and outcome control are not replicated across platforms. 

\subsubsection{Average Absolute Deviation}
Next, we use a similar specification controlling additionally for the observed model error, to study the effects of process and outcome controls on the average absolute deviation and control. Different from the stated model error which was shown to the participants before they decided if they would want to use the models, the observed model error is defined as the mean average absolute error of the subset of twenty students the participants made predictions for. Prior work shows that people’s trust in a model is affected by both its stated accuracy and its observed accuracy, and that the effect of stated accuracy can change depending on the observed accuracy \cite{ying2019understanding}. Given the length of the exercise, it is possible that participants may form judgments about the models based on experienced discrepancy between models' and their own predictions. 

We first discuss the average treatment effects of outcome and process control on average absolute deviation for all participants regardless of whether they used the models or not, which were previously summarized in Figure ~\ref{fig:all_study_summary} . For participants who chose not to use the models, the deviation is still calculated as the difference between their predictions and the models' predictions, even though they did not see the models' predictions during the tasks.

We use the following regression model, where $\beta_9$ is a vector of coefficients for the demographic variables:

{\begin{align} \label{eq:2}  
& \text{AAD}_{i} = \beta_0 + \beta_1\cdot\text{stated\_model\_error}_i  \\ \nonumber & + \beta_2\cdot\text{observed\_model\_error}_i  +  \beta_3\cdot\text{change\_by\_10}_i  \\ \nonumber & + 
\beta_4\cdot\text{change\_freely}_{i} + \beta_5\cdot\text{change\_input}_{i}  \\ \nonumber & + 
\beta_6\cdot\text{change\_algorithm}_{i}  + \beta_7\cdot\text{(change\_input * change\_freely)}_{i}  \\
\nonumber & + \beta_8\cdot\text{(change\_algorithm * change\_freely)}_{i} + \beta_9\cdot\text{demographics} + \epsilon_{i}
\end{align}}

The regression results are summarized in Table~\ref{table:deviation_unfiltered}. We find that neither the stated nor the observed model performance affects how much the participants would deviate from the models. Allowing participants to change the outcome by a limited amount appears to decrease the deviation while allowing them to change the outcome freely appears to increase the deviation. Changing algorithms rather than changing input factors statistically significantly reduces participants' deviations from models' predictions on MTurk and Prolific respectively. However, this effect is again canceled out by the interaction effect of being able change both the predictions and the algorithms. 

\begin{table*}[htbp]\centering
\def\sym#1{\ifmmode^{#1}\else\(^{#1}\)\fi}
\caption{Intervention Effects on Average Absolute Deviation}\small
\begin{tabular}{l*{6}{c}}
\hline\hline
            &\multicolumn{1}{c}{(1)}&\multicolumn{1}{c}{(2)}&\multicolumn{1}{c}{(3)}&\multicolumn{1}{c}{(4)}&\multicolumn{1}{c}{(5)}&\multicolumn{1}{c}{(6)}\\
            &\multicolumn{1}{c}{MTurk}&\multicolumn{1}{c}{MTurk}&\multicolumn{1}{c}{Prolific}&\multicolumn{1}{c}{Prolific }&\multicolumn{1}{c}{Pooled}&\multicolumn{1}{c}{Pooled}\\
\hline
stated model error&       0.041         &       0.222         &      -0.115         &      -0.014         &       0.039         &       0.074         \\
            &     (0.396)         &     (0.391)         &     (0.406)         &     (0.417)         &     (0.283)         &     (0.283)         \\
[1em]
observed model error&       0.001         &       0.002         &       0.040         &       0.040         &       0.024         &       0.032         \\
            &     (0.083)         &     (0.083)         &     (0.097)         &     (0.097)         &     (0.064)         &     (0.064)         \\
[1em]
change outcome by 10&      -1.530         &      -1.295         &             &               &      -2.258\sym{*}  &      -1.954         \\
            &     (1.169)         &     (1.184)         &              &             &     (1.014)         &     (1.018)         \\
[1em]
change outcome freely&       1.194         &       1.080         &       3.388\sym{*}  &       3.223\sym{*}  &       2.282\sym{**} &       2.291\sym{**} \\
            &     (1.074)         &     (1.081)         &     (1.333)         &     (1.323)         &     (0.839)         &     (0.835)         \\
[1em]
change input&       0.247         &       0.365         &       0.536         &       0.366         &       0.425         &       0.572         \\
            &     (1.720)         &     (1.720)         &     (1.955)         &     (1.954)         &     (1.288)         &     (1.279)         \\
[1em]
change algorithm &      -4.980\sym{***}&      -4.753\sym{***}&      -4.884\sym{**} &      -5.041\sym{**} &      -4.894\sym{***}&      -4.766\sym{***}\\
            &     (1.309)         &     (1.312)         &     (1.700)         &     (1.695)         &     (1.047)         &     (1.048)         \\
[1em]
change outcome and input &       1.897         &       1.876         &      -0.382         &      -0.276         &       0.785         &       0.565         \\
            &     (1.787)         &     (1.791)         &     (1.970)         &     (1.966)         &     (1.320)         &     (1.311)         \\
[1em]
change outcome and algorithm &       5.196\sym{***}&       5.027\sym{***}&       2.654         &       2.809         &       4.037\sym{***}&       3.846\sym{***}\\
            &     (1.508)         &     (1.503)         &     (1.832)         &     (1.825)         &     (1.168)         &     (1.167)         \\
[1em]
demographics &               &     \cmark  &               &     \cmark         &              &      \cmark        \\
[1em]
\_cons      &       9.222         &      -0.842         &      13.310         &       9.817         &       9.554         &       6.111         \\
            &     (7.734)         &     (8.476)         &     (7.812)         &     (8.622)         &     (5.495)         &     (6.084)         \\
\hline
\(N\)       &        1,443         &        1,443         &        1,172         &        1,172         &        2,615         &        2,615         \\
\(R^{2}\)   &       0.044         &       0.084         &       0.052         &       0.082         &       0.050         &       0.074         \\
adj. \(R^{2}\)&       0.038         &       0.063         &       0.046         &       0.057         &       0.047         &       0.062         \\
\hline\hline
\multicolumn{7}{l}{\footnotesize Prolific specification excludes change outcome by 10 because this condition was not replicated on Prolific}\\
\multicolumn{7}{l}{\footnotesize The sample includes participants who are from adjust-by-10}\\
\multicolumn{7}{l}{\footnotesize Standard errors in parentheses}\\
\multicolumn{7}{l}{\footnotesize \sym{*} \(p<0.05\), \sym{**} \(p<0.01\), \sym{***} \(p<0.001\)}\\
\end{tabular}
\label{table:deviation_unfiltered}
\end{table*}

\begin{figure}[h]
\includegraphics[scale=0.4]
{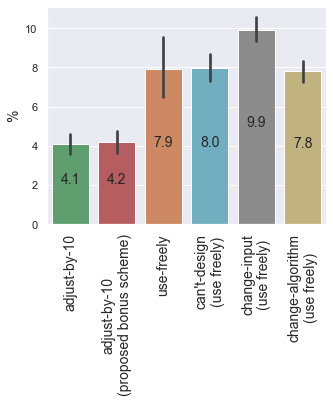}
\caption{Average Absolute Deviations for Participants Who Can Change Models' Predictions and Chose to Use the Models. Participants in the adjust-by-10 conditions deviated the least from the models among all groups. Other groups that can change models' predictions freely deviated similarly from the models.}
\Description{Average Absolute Deviations for Participants Who Can Change Models' Predictions and Chose to Use the Models. Participants in the adjust-by-10 conditions deviated the least from the models among all groups. Other groups that can change models' predictions freely deviated similarly from the models.}
\label{fig:deviation_subgroup}
\end{figure}

Additionally, we want to understand for the participants who were able to change the models' predictions and who chose to use the models (so they were able to see the models' predictions), how much they deviated from the models' predictions. We see that as illustrated in Figure ~\ref{fig:deviation_subgroup}, unsurprisingly participants who could change the predictions by a limited amount on average deviated significantly less from others by almost a half. The use-freely, can't-design (use freely), and change-algorithm (use freely) groups deviated similarly from models' predictions by around 8 percentile points. The change-input (use freely) group, however, deviated the most by 9.9 percentile points. This shows that unless participants are only allowed to change predictions restrictively, they deviate similarly from models' predictions under outcome and process control. 

\subsubsection{Average Absolute Errors}
\label{section: comparison_AAE}
Lastly, we are interested in how outcome and process controls affect users' AAE as defined in \textsection\ref{section:RQs} and Table ~\ref{table:regression_notation}. For reasons discussed previously, we use the same specification accounting for stated and observed errors in analyzing AAE, where $\beta_9$ is a vector of coefficients for the demographic variables: 

{\begin{align}\label{eq:3}  
& \text{AAE}_{i} = \beta_0 + \beta_1\cdot\text{stated\_model\_error}_i \\\nonumber 
& + \beta_2\cdot\text{observed\_model\_error}_i + \beta_3\cdot\text{change\_by\_10}_i \\\nonumber 
& + \beta_4\cdot\text{change\_freely}_{i}
+ \beta_5\cdot\text{change\_input}_{i} \\\nonumber 
& + \beta_6\cdot\text{change\_algorithm}_{i}  + \beta_7\cdot\text{(change\_input * change\_freely)}_{i} 
\\\nonumber 
& + \beta_8\cdot\text{(change\_algorithm * change\_freely)}_{i} + \beta_9\cdot\text{demographics} + \epsilon_{i}
\end{align}}

\begin{table*}[htbp]\centering
\def\sym#1{\ifmmode^{#1}\else\(^{#1}\)\fi}\small
\caption{Intervention Effects on AAE}
\begin{tabular}{l*{6}{c}}
\hline\hline
            &\multicolumn{1}{c}{(1)}&\multicolumn{1}{c}{(2)}&\multicolumn{1}{c}{(3)}&\multicolumn{1}{c}{(4)}&\multicolumn{1}{c}{(5)}&\multicolumn{1}{c}{(6)}\\
            &\multicolumn{1}{c}{MTurk}&\multicolumn{1}{c}{MTurk}&\multicolumn{1}{c}{Prolific}&\multicolumn{1}{c}{Prolific}&\multicolumn{1}{c}{Pooled}&\multicolumn{1}{c}{Pooled}\\
\hline
stated model error&      -0.083         &      -0.019         &      -0.327         &      -0.271         &      -0.189         &      -0.162         \\
            &     (0.221)         &     (0.220)         &     (0.230)         &     (0.235)         &     (0.159)         &     (0.160)         \\
[1em]
observed model error &       0.727\sym{***}&       0.730\sym{***}&       0.678\sym{***}&       0.684\sym{***}&       0.706\sym{***}&       0.713\sym{***}\\
            &     (0.046)         &     (0.046)         &     (0.054)         &     (0.055)         &     (0.035)         &     (0.035)         \\
[1em]
change outcome by 10&      -1.601\sym{*}  &      -1.452\sym{*}  &           &               &      -1.927\sym{***}&      -1.838\sym{***}\\
            &     (0.623)         &     (0.630)         &                 &                 &     (0.555)         &     (0.557)         \\
[1em]
change outcome freely&      -0.396         &      -0.419         &       0.119         &       0.079         &      -0.125         &      -0.120         \\
            &     (0.576)         &     (0.573)         &     (0.724)         &     (0.722)         &     (0.450)         &     (0.449)         \\
[1em]
change input&       0.871         &       1.028         &       0.590         &       0.488         &       0.758         &       0.807         \\
            &     (0.881)         &     (0.885)         &     (1.026)         &     (1.026)         &     (0.664)         &     (0.660)         \\
[1em]
change algorithm &      -2.112\sym{***}&      -1.934\sym{**} &      -2.255\sym{**} &      -2.377\sym{**} &      -2.161\sym{***}&      -2.107\sym{***}\\
            &     (0.637)         &     (0.638)         &     (0.854)         &     (0.858)         &     (0.515)         &     (0.519)         \\
[1em]
change outcome and input &      -0.007         &      -0.122         &       0.596         &       0.618         &       0.302         &       0.206         \\
            &     (0.921)         &     (0.928)         &     (1.034)         &     (1.039)         &     (0.684)         &     (0.682)         \\
[1em]
change outcome and algorithm&       2.641\sym{***}&       2.498\sym{**} &       1.652         &       1.725         &       2.169\sym{***}&       2.091\sym{***}\\
            &     (0.796)         &     (0.792)         &     (0.943)         &     (0.949)         &     (0.604)         &     (0.607)         \\
[1em]
demographics &               &     \cmark  &               &     \cmark         &              &      \cmark        \\
[1em]
\_cons      &      11.320\sym{**} &       7.540         &      17.886\sym{***}&      15.718\sym{**} &      14.132\sym{***}&      12.383\sym{***}\\
            &     (4.320)         &     (4.932)         &     (4.448)         &     (5.136)         &     (3.081)         &     (3.552)         \\
\hline
\(N\)       &        1,443         &        1,443         &        1,172         &        1,172         &        2,615         &        2,615         \\
\(R^{2}\)   &       0.163         &       0.193         &       0.148         &       0.168         &       0.157         &       0.174         \\
adj. \(R^{2}\)&       0.158         &       0.175         &       0.143         &       0.144         &       0.155         &       0.164         \\

\hline\hline
\multicolumn{7}{l}{\footnotesize Prolific specification excludes change outcome by 10 because this condition was not replicated on Prolific}\\
\multicolumn{7}{l}{\footnotesize The sample includes participants who are from adjust-by-10}\\
\multicolumn{7}{l}{\footnotesize (proposed bonus scheme)}\\
\multicolumn{7}{l}{\footnotesize Standard errors in parentheses}\\
\multicolumn{7}{l}{\footnotesize \sym{*} \(p<0.05\), \sym{**} \(p<0.01\), \sym{***} \(p<0.001\)}\\
\end{tabular}
\label{table:outcome_aae_unfiltered}
\end{table*}

The results are summarized in Table ~\ref{table:outcome_aae_unfiltered}. Again, we observe that the stated model error rate is not significantly associated with participants' AAE. However, the observed model error is highly associated with users' AAE. This is not surprising because, for participants who chose to use the models and cannot change the models' predictions, the observed error rate is the same as participants' AAE.

We observe that changing outcome by a limited amount statistically significantly reduces participants' errors by 1.6 percentile points (Mturk column(1)). Furthermore, changing algorithm is shown to statistically significantly reduce participants' errors by 2.1 (column(1))and 2.3 (column (3)) percentile points respectively on MTurk and Prolific. These results are likely more significant compared to in Figure ~\ref{fig:all_study_summary} because we averaged the identical conditions in study 1 and 2. However, if participants can change both the algorithm and the outcome, the additional effects of doing so would increase participants' prediction errors which cancel out the reduction in the error from just changing the algorithm alone, resulting in similar prediction performances among participants who could change the algorithm alone and who would change both the outcome and algorithm. 

Additionally, we want to understand how choosing to use models and reliance on the models' predictions are associated with prediction performance. We already see in Figure ~\ref{fig:study1_error_distribution_summary} \textsection\ref{section:dietvorst_replication} that choosing to use the model is associated with lower AAEs and prevents participants from making large errors. We observe similar results for all other groups as illustrated by Figure ~\ref{fig:study23_error_distribution_summary} in Appendix ~\ref{appendix:supplement_figures}. 

\begin{figure}[h]
\includegraphics[width=0.5\textwidth, scale=0.5]
{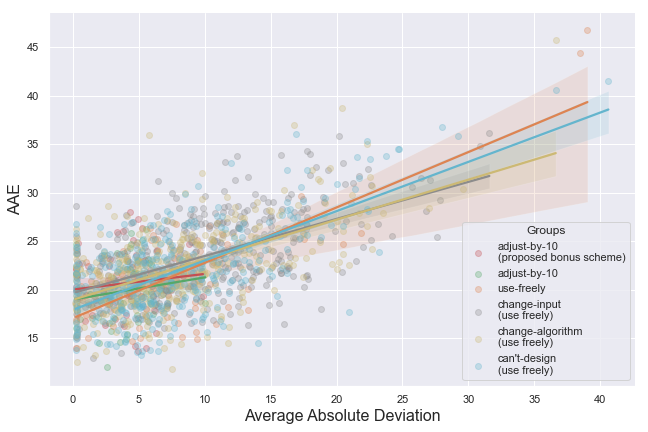}
\caption{Average Absolute Errors VS. Average Absolute Deviations for Participants Who Chose to Use the Models and Could Modify Models' Predictions. Larger deviation from the models' predictions are associated with higher errors.}
\Description{Average Absolute Errors VS. Average Absolute Deviations for Participants Who Chose to Use the Models and Could Modify Models' Predictions. Larger deviation from the models' predictions are associated with higher errors.}
\label{fig:aae_subgroup}
\end{figure}

Lastly, for participants who chose to use the models and could change models' predictions, we observe that higher deviation from the models' predictions is associated with higher AAEs in Figure~\ref{fig:aae_subgroup}. Thus, both choosing to use models and deviating less from models' predictions are associated with lower prediction errors for competing our particular tasks. However, it is important to note that in many real-life cases, experienced workers are capable of identifying and correcting for algorithmic system failures to achieve better performances \cite{de2020case,cheng2022child}. 

\newpage
\section{Supplement Figures}
\label{appendix:supplement_figures}

\begin{figure}[h]
\includegraphics[width=0.5\textwidth, scale=0.5]
{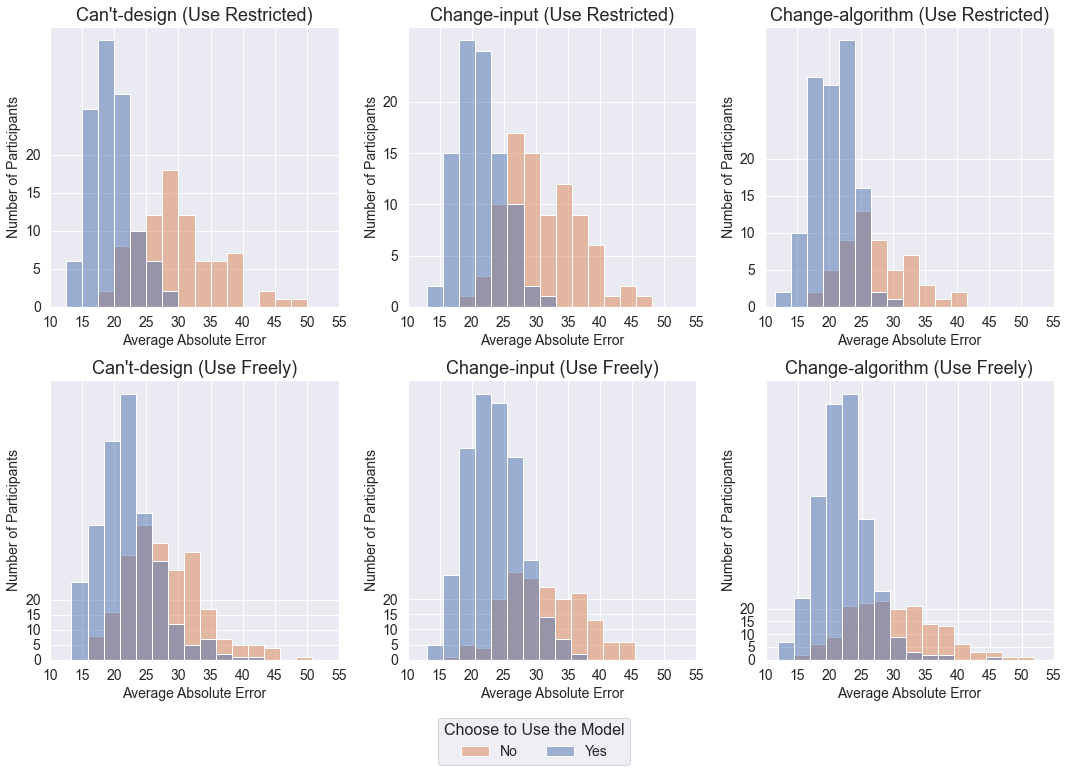}
\caption{Distribution of Errors for Participants in Study 2 and 3. The distributions of AAEs by participants who chose to use models for the tasks appear left-skewed, suggesting they are less likely to make large errors.}
\Description{Distribution of Errors for Participants in Study 2 and 3. The distributions of AAEs by participants who chose to use models for the tasks appear left-skewed, suggesting they are less likely to make large errors.}
\label{fig:study23_error_distribution_summary}
\end{figure} 

\begin{figure}[h]
\includegraphics[width=0.5\textwidth, scale=0.5]
{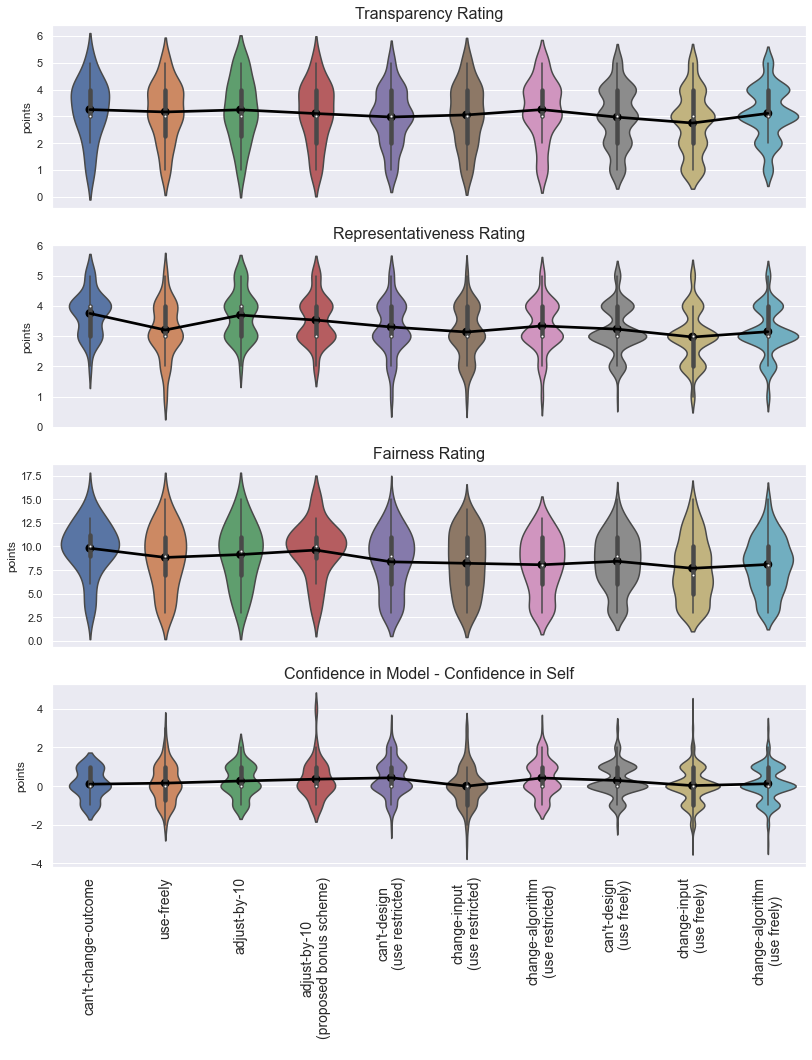}
\caption{Ratings on Transparency, Representativensss, Fairness, and Differences in Confidence in the Model and Oneself for Each Experimental Condition. There are no significant differences on these ratings across conditions.}
\Description{Ratings on Transparency, Representativensss, Fairness, and Differences in Confidence in the Model and Oneself for Each Experimental Condition. There are no significant differences on these ratings across conditions.}
\label{fig:trf_ratings}
\end{figure} 

\newpage
\section{Results on Participants Who Passed Attention Checks}
\label{appendix:filtered_results}
As discussed in \textsection\ref{section:participants_description}, we focus on presenting results using the full data including participants who did not pass the attention checks (i.e. who failed to choose the correct statements related to "percentile" either at the beginning or the end of the survey) because there is no detectable differences among the groups. 

We present the findings on the subset of participants who passed the attention checks here and show that filtering on attention checks does not qualitatively change our findings. 

There are 509 participants who did not pass either attention check. Table ~\ref{table:demographics_all_filtered} summarizes the demographics characteristics of the participants who passed the attention checks, which is similar to the characteristics of the full sample.

\begin{table}[h]\centering\small
\def\sym#1{\ifmmode^{#1}\else\(^{#1}\)\fi}
\caption{Demographics of Participants from MTurk, Prolific and Both (Pooled)}
\begin{tabular}{p{5.8cm}p{0.6cm}p{0.6cm}p{0.6cm}}
& Mturk & Prolific & Pooled \\
\hline\hline
\% Female                                                               & 40.3  & 66.0     & 54.0   \\ [0.02cm]
\% Male                                                                & 56.6  & 27.9     & 41.3   \\ [0.02cm]
\% White                                                                 & 81.9  & 74.4     & 77.9   \\ [0.02cm]
\% Black or African American                                             & 8.2   & 6.8      & 7.4    \\ [0.02cm]
\% Hispanic or Latino                                                    & 8.9  & 10.9     & 10.0   \\ [0.02cm]
Average Age (Year)                                                           & 38.2  & 34.3     & 36.2   \\ [0.02cm]
\% Some college and above                                                & 92.2  & 88.1     & 90.0   \\ [0.02cm]
\% High school/GED                                                       & 7.3   & 10.8     & 9.2    \\ [0.02cm]
Average Confidence in Math            & 3.2  & 2.7     & 2.9   \\ [0.02cm]
\% Participated in Algorithm-related Studies Before & 76.1  & 50.8       & 62.6     \\ [0.02cm]
\hline
N   & 940 & 1,078     & 2,018 \\
\hline\hline
\multicolumn{4}{l}{\footnotesize This sample does not include adjust-by-10 (proposed bonus scheme) described in }\\
\multicolumn{4}{l}{\footnotesize Appendix~\ref{appendix:adjust_by_10_alt}.}\\
\multicolumn{4}{l}{\footnotesize Response categories where few respondents selected or declined to answer are dropped.}\\
\multicolumn{4}{l}{\footnotesize Average confidence in math is converted from a survey question using a 5-point }\\
\multicolumn{4}{l}{\footnotesize Likert scale where 1 is "not confident" and 5 is "extremely confident."}\\
\end{tabular}
\label{table:demographics_all_filtered}
\end{table}

Figure~\ref{fig:study1_filtered} summarizes the replication results from Study 1. All study conditions' results for the sub-population are summarized in Figure~\ref{fig:total_bar_filtered}.

\subsection{Study1}
\begin{figure}[h]
\includegraphics[scale=0.3]{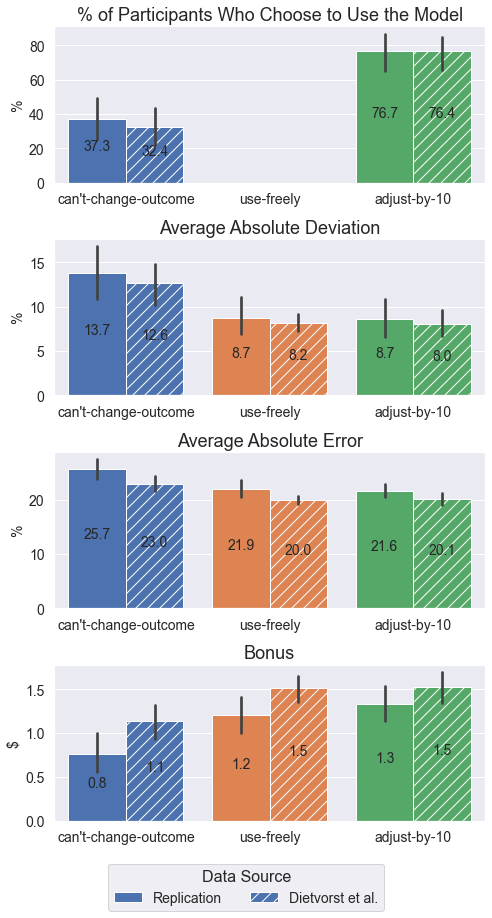}
\caption{Comparing Findings From Replication and Dietvorst et al (Participants Who Passed Attention Checks). The non-shaded bars and shaded bars are results from our replication and Dietvorst et al.'s study respectively. We show that filtering out participants who did not pass the attention checks does not change our conclusion about study 1. We still replicate the results from Dietvorst et al.}
\Description{Comparing Findings From Replication and Dietvorst et al (Participants Who Passed Attention Checks). The non-shaded bars and shaded bars are results from our replication and Dietvorst et al.'s study respectively. We show that filtering out participants who did not pass the attention checks does not change our conclusion about study 1. We still replicate the results from Dietvorst et al.}
\label{fig:study1_filtered}
\end{figure}

As shown in Figure~\ref{fig:study1_filtered}, filtering out participants who did not pass the attention checks does not change our conclusion about study 1 that we successfully replicate the results from Dietvorst et al.

As discussed in the main text, we note an important caveat to the replication of \cite{Dietvorst18} namely that for the same treatment where participants cannot modify the model nor its predictions, we observed an unexpectedly high take-up rate in (69\%, Study 2, can't-design (use restricted)) compared to (37.3\%, Study 1, can't-change-outcome) when we expected the take-up rates to be the same as shown in Figure~\ref{fig:total_bar_filtered}. Thus, if we compare across studies and use (69\%, Study 2, can't-design (use restricted)) as the reference, we observe no statistically significant difference between it and (76.7\%, Study 1, adjust-by-10) in their take-up rates. Additionally we observe no difference in the take-up rates between (69\%, Study 2 can't-design (use restricted)) and (69.3\%, Study 3 can't-design (use freely)) when we expect the later group to have a higher take-up rate if the conclusion about giving participants the ability to change the outcome in \cite{Dietvorst18} holds. Therefore, the caveat that cohort effects complicate the interpretation of the replication remains when we look at the subset of the participants. 

\subsection{Study2}
\label{appendix:subsample_study2}

\begin{figure*}[h]
\includegraphics[scale=0.4]{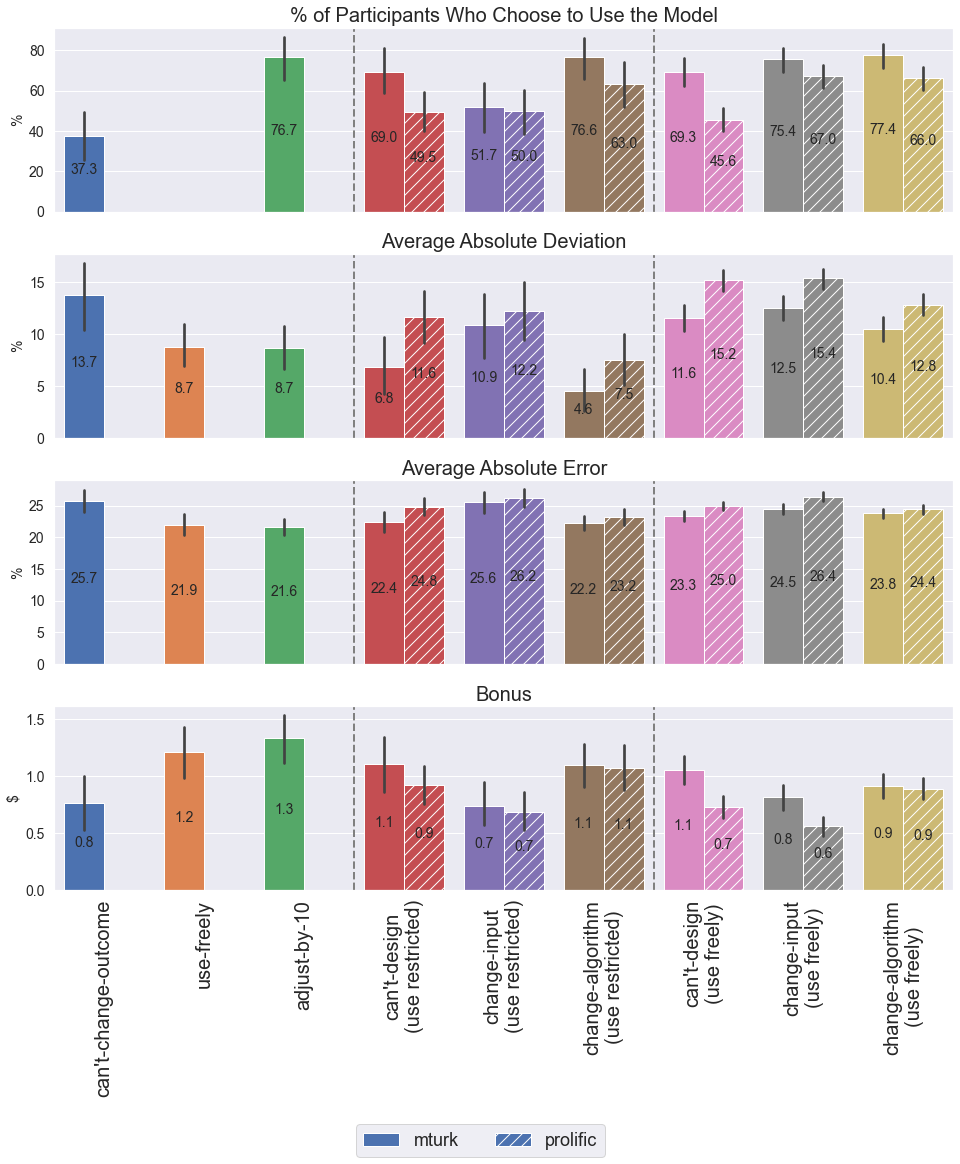}
\caption{All Experimental Conditions Results on MTurk and Prolific Platforms (Participants Who Passed Attention Checks). The gray dotted vertical line separates each study. The left, middle, and right frames correspond to study 1, 2 and 3 respectively. Use-freely group received models without having to choose to use the models or not, which means 100\% of the participants chose to use the model by default and the corresponding bar is removed in the first row.}
\Description{All Experimental Conditions Results on MTurk and Prolific Platforms (Participants Who Passed Attention Checks). The gray dotted vertical line separates each study. The left, middle, and right frames correspond to study 1, 2 and 3 respectively. Use-freely group received models without having to choose to use the models or not, which means 100\% of the participants chose to use the model by default and the corresponding bar is removed in the first row.}
\label{fig:total_bar_filtered}
\end{figure*}

Overall, again we observe large differences between the same conditions across batches and across platforms in Figure~\ref{fig:total_bar_filtered}.

With regard to process control in the form of changing input, we observe again that changing input does not increase \% of participants who chose to use the model according to results on both MTurk and Prolific. For participants in Study 2, given that they could not change the models predictions, participants who could change the models' inputs (51.7\%, Study 2, change-input (use restricted)) are not more likely to chose the models compared to the ones who could not modify the model (69\%, Study 2, can't-design (use restricted)). 

Similarly, participants who could choose the training algorithm (76.6\%, Study 2, change-algorithm (use restricted)) are more likely to use the model than the ones who could not modify the models (69\%, Study 2, can't-design (use restricted)) on MTurk. This observation holds qualitatively for Prolific participants as well, although the difference between (49.5\%, Study 2, can't-design (use restricted)) and (63.6\%, Study 2, change-algorithm (use restricted)) appears less statistically significant ($\chi^2(1, N=176)=3.15, p = 0.076$) compared to the results on the full sample. 

\subsection{Study 3}
On MTurk, we observe no compounding effects of administering both process and outcome controls for participants who could change the training procedure: the take-up rates are indistinguishable between (76.7\%, Study 1, adjust-by-10), 76.6\%, change-algorithm (use restricted), change-algorithm (use freely). This means that allowing participants to change the outcome mitigates aversion to the same extent as allowing them to change the training procedure, and allowing them to change both the outcome and training procedure does not further mitigate aversion. 

As for changing-input, there is no statistically significant effects of changing input when participants had no outcome control (69\%, Study 2, can't-design (use restricted) vs. 51.7\%, Study 2, change-input (use restricted)) or when they had outcome control (69.3\%, Study 3, can't-design (use freely) vs. 75.4\%, Study 3, change-input (use freely)). Together, this shows that when participants had outcome control or could change the training algorithm, \textit{or both} they were more likely to choose to use the model compared to if they have no control at all. This aligns with the finding in the main text. 

On Prolific, we do not observe effect of outcome control alone on mitigating algorithm aversion as mentioned in \textsection\ref{appendix:subsample_study2}, therefore our conclusions about the compounded effects of outcome and process control differ from MTurk. In study 3 where participants have outcome control, they are significantly ($\chi^2(1, N=551)= 25.8, p < 0.000$) more likely to choose to use the model when they can change model inputs (67\%, Study 3, change-input (use freely)) compared to when they cannot (46.6\%, Study 3, can't-design (use freely)). Similarly, when we focus on participants can change model inputs, they are significantly ($\chi^2(1, N=356) = 8.11, p = 0.004$) more likely to choose to use the model when provided outcome control (67\%, Study 3, change-input (use freely)) compared to when they have no outcome control (50\%, Study 2, change-input (use restricted)). Therefore, we observe a significant interaction effect between process control in the form of changing inputs and outcome control in our experiments on the Prolific platform. Whereas changing inputs alone or being able to deviate freely from model predictions on its own does not mitigate algorithm aversion, having both forms of control has a strong and statistically significant mitigating effect. This again aligns with our conclusion in the main text. 
\end{document}